\documentclass[twocolumn]{aastex631}

\usepackage{amsmath, nccmath}
\usepackage{natbib}
\usepackage{multirow}
\usepackage{float}
\usepackage{ulem}
\usepackage{xcolor}

\shorttitle{Near-IR Spectral Properties of SNe Ib/Ic Progenitors}
\shortauthors{Jung \& Yoon}

\graphicspath{{./}}

\begin{document}

\title{Near-infrared Spectral Properties of Type Ib/Ic Supernova Progenitors \\
and Implications for JWST and NGRST Observations}

\author[0000-0002-8074-6616]{Moo-Keon Jung}
\affiliation{Department of Physics and Astronomy, Seoul National University, Seoul 08826, Korea}

\author[0000-0002-5847-8096]{Sung-Chul Yoon}
\affiliation{Department of Physics and Astronomy, Seoul National University, Seoul 08826, Korea}
\affiliation{SNU Astronomy Research Center, Seoul National University, Seoul 08826, Korea}

\defcitealias{Yoon2017MNRAS.470.3970Y}{Y17}
\defcitealias{Yoon2019ApJ...872..174Y}{Y19}
\defcitealias{Jung2022ApJ...925..216J}{JYK22}

\begin{abstract}
While about 20 Type II supernova progenitors have been identified using optical data from the Hubble Space Telescope (HST), direct detection of type Ib/Ic supernova (SN Ib/Ic) progenitors remains challenging due to their faint optical brightness and highly obscured environments. This study aims to investigate the detection limits and advantages of near-infrared (near-IR) observations with the James Webb Space Telescope (JWST) and the Nancy Grace Roman Space Telescope (NGRST) for the detection of SN Ib/Ic progenitors. The spectral energy distributions of SN Ib/Ic progenitor models with various masses, chemical compositions, and mass-loss rates are calculated with the non-LTE radiative transfer code CMFGEN. We then assess the detectability of SN Ib/Ic progenitors using near-IR filters from the JWST and the NGRST, comparing the results to the capabilities of the HST. Our analysis indicates that near-IR observations significantly outperform the HST in detecting SN Ib/Ic progenitors when considering the effect of extinction. Near-IR magnitudes also provide better constraints on the mass-loss rates of progenitors because of the free-free emission from the wind matter. Additionally, near-IR magnitudes and color-color diagrams are effective in distinguishing SN Ib/Ic progenitors from possible companion and/or background objects. This study suggests that the JWST and the NGRST can play a crucial role in advancing our understanding of SN Ib/Ic progenitors by improving detectability and offering better constraints on progenitor properties. We emphasize that observations with exposure times exceeding 1 hour would be needed to detect typical SNe Ib/Ic progenitors at distances greater than 10 Mpc.

\end{abstract}

\section{Introduction} \label{sec:intro}

Core-collapse supernovae, including Type Ib, Ic, and II, represent the explosive deaths of massive stars (typically $M_\mathrm{init} > 8 M_\odot$), driven by the collapse of their cores. Type Ib and Ic supernovae (SNe Ib/Ic) lack hydrogen lines in their spectra \citep{Filippenko1997ARA&A..35..309F}, indicating that the hydrogen envelopes of their progenitors are stripped off during stellar evolution. The progenitor scenario for SNe Ib/Ic is a topic of debate. Traditionally proposed progenitors are massive Wolf-Rayet (W-R) stars that have lost their hydrogen envelope by wind mass-loss, which have relatively high zero-age main sequence (ZAMS) mass of $M_\mathrm{ZAMS} \gtrsim 20-35M_\odot$. An alternative scenario suggests hydrogen-deficient progenitors produced through close binary interactions \citep[see][for a review]{Yoon2015PASA...32...15Y}.

The Hubble Space Telescope (HST) has played a crucial role in the direct detection of supernova (SN) progenitors. Prior to the HST, SN studies primarily relied on follow-up observations due to the unpredictability of SN explosions, which typically occur at vast distances. The advent of the HST enabled deep pre-explosion imaging, leading to the direct identification of dozens of type II supernova (SN II) progenitors in archival data \citep[e.g.,][]{VanDyk2003PASP..115.1289V,  Smartt2004Sci...303..499S, Maund2005MNRAS.364L..33M, Hendry2006MNRAS.369.1303H, Li2006ApJ...641.1060L, Li2007ApJ...661.1013L, Smartt2009ARA&A..47...63S,  Elias-Rosa2010ApJ...714L.254E, Elias-Rosa2011ApJ...742....6E, Fraser2011MNRAS.417.1417F, VanDyk2012ApJ...756..131V, Maund2013MNRAS.431L.102M, Fraser2014MNRAS.439L..56F, Smartt2015PASA...32...16S, Kochanek2017MNRAS.467.3347K, Tartaglia2017ApJ...836L..12T, Kilpatrick2018MNRAS.481.2536K, O'Neil2019A&A...622L...1O, Kilpatrick2023ApJ...952L..23K,  Pledger2023ApJ...953L..14P, Xiang2024ApJ...969L..15X}. These direct detections provide valuable insights into the late-time evolution of massive stars.

However, detection of progenitors of SNe Ib/Ic remains challenging. To date, only one progenitor of a SN Ib, iPTF13bvn, has been confirmed \citep{Cao2013ApJ...775L...7C, Eldridge2016MNRAS.461L.117E, Folatelli2016ApJ...825L..22F}. Recently, another SN Ib progenitor candidate for SN 2019yvr has been discovered \citep{Kilpatrick2021MNRAS.504.2073K}. \citet{Kilpatrick2018MNRAS.480.2072K} and \citet{VanDyk2018ApJ...860...90V} reported a potential progenitor candidate for SN Ic 2017ein, but a more recent study by \citet{Zhao2025ApJ...980L...6Z} concluded that it was not the SN progenitor. Most progenitors of SNe Ib/Ic have not been directly identified in pre-explosion images from ground-based telescopes and the HST \citep{Eldridge2013MNRAS.436..774E, Strotjohann2024ApJ...960...72S}, implying their faintness in optical wavelengths. While even very massive SN Ib/Ic progenitors \citep[$M > 9M_\odot$ corresponding to $M_\mathrm{ZAMS} \gtrsim 40M_\odot$;][]{Yoon2017MNRAS.470.3970Y, Woosley2019ApJ...878...49W} might be optically faint \citep{Yoon2012A&A...544L..11Y}, recent studies on SN light curves suggest that most SN Ib/Ic progenitors have masses below 10 $M_\odot$ \citep{Lyman2016MNRAS.457..328L, Gangopadhyay2020MNRAS.497.3770G, Pandey2021MNRAS.507.1229P}, and the estimated fraction of progenitors with $M_\mathrm{ZAMS} > 25M_\odot$ seem to be limited to about 10\% \citep{Karamehmetoglu2023A&A...678A..87K}. Relatively low-mass progenitors, which constitute the majority, support the binary-stripped progenitor scenario \citep[e.g.,][]{Podsiadlowski1992ApJ...391..246P, Woosley1995ApJ...448..315W, Vanbeveren1998NewA....3..443V, Wellstein1999A&A...350..148W, Wellstein2001A&A...369..939W, Eldridge2004MNRAS.353...87E, Richardson2006AJ....131.2233R, Eldridge2008MNRAS.384.1109E, Yoon2010ApJ...725..940Y, Dessart2011MNRAS.414.2985D,  Drout2011ApJ...741...97D, Cano2013MNRAS.434.1098C, Bianco2014ApJS..213...19B, Modjaz2014AJ....147...99M, Dessart2015MNRAS.453.2189D, Taddia2015A&A...574A..60T, Yoon2015PASA...32...15Y, Liu2016ApJ...827...90L, Prentice2016MNRAS.458.2973P, Yoon2017ApJ...840...10Y, Taddia2018A&A...609A.136T, Gilkis2019MNRAS.486.4451G, Prentice2019MNRAS.485.1559P, Dessart2020A&A...642A.106D, Laplace2020A&A...637A...6L, Barbarino2021A&A...651A..81B, Laplace2021A&A...656A..58L, Schneider2021A&A...645A...5S, Woosley2021ApJ...913..145W, Aguilera-Dena2022A&A...661A..60A, Zheng2022MNRAS.512.3195Z, Aguilera-Dena2023A&A...671A.134A, Dessart2024A&A...685A.169D, Ercolino2024A&A...685A..58E}.

A stripped-envelope progenitor reveals a hot and compact core at the surface, leading to the prediction that the optical magnitude of a SN Ic progenitor is only $M_v$\footnote{$M_v$ denotes the absolute magnitude of $v$ intermediate-band filter (3900\AA $\lesssim \lambda \lesssim$ 4500\AA) as defined in \citet{Stromgren1956VA......2.1336S}.} $\sim -1.5$ to $-2.5$ under the blackbody assumption \citep{Yoon2012A&A...544L..11Y}. The predicted optical brightness of the SN Ic progenitors is markedly lower than that of the SN Ib progenitors ($M_v \simeq -4$) when the effects of stellar winds are disregarded. However, the presence of optically thick winds can significantly enhance the optical and infrared brightness of the progenitors by forming photospheres far from the stellar surface at considerably lower effective temperatures compared to the surface temperatures. The optically thick winds also contribute to strong free-free emission and emission lines \citep[see][]{Jung2022ApJ...925..216J}.

The optical magnitudes of both SN Ib and Ic progenitors are adjusted to $M_\mathrm{F555W}$ $\sim -4.5$ to $-5.0$ when considering the standard wind mass-loss rate. They could be detected with deep HST observations with exposure time ($t_\mathrm{exp}$) exceeding 1 hour for SN Ib/Ic progenitors located at a distance of $d=20$ Mpc. However, most HST archival images have been too shallow ($t_\mathrm{exp} \lesssim 20$ min) to directly detect the progenitors. In addition to deeper HST observations, the possibility of direct detection with near-infrared (near-IR) instruments can also be considered as stellar winds, especially free-free emission, can significantly increase the near-IR flux of the progenitors.

In this study, we investigate the detection possibility and advantages of using near-IR observations from the James Webb Space Telescope (JWST) and the Nancy Grace Roman Space Telescope (NGRST) to identify SN Ib/Ic progenitors. The non-LTE radiative transfer code CMFGEN \citep{Hillier1998ApJ...496..407H} is employed to calculate the spectral energy distributions (SEDs) of SN Ib/Ic progenitors. In Section \ref{sec:model}, we describe the SN Ib/Ic progenitor models and their assumed wind properties. In Section \ref{sec:results}, we present the calculated SEDs of the SN Ib/Ic progenitor models and discuss the effects of winds on the near-IR magnitudes. In Section \ref{sec:detectability}, we compare the detectability of the HST, JWST, and NGRST for SN Ib/Ic progenitors, accounting for extinction effects. In Section \ref{sec:advantage}, we highlight the information provided by near-IR magnitudes and colors, and discuss their advantages over optical data. We conclude and summarize our study in Section \ref{sec:conclusions}.

\section{Models} \label{sec:model}

The SN Ib/Ic progenitor models and their atmospheric models used in this study are identical to those in \citet[][hereafter \citetalias{Jung2022ApJ...925..216J}]{Jung2022ApJ...925..216J}. Therefore, we provide only the key information here. For further details on the models, please refer to Section 2 of \citetalias{Jung2022ApJ...925..216J} and the references therein.

\subsection{Input Progenitor Models} \label{sec:input}

\begin{deluxetable*}{lrccrrcrccccc}
\tablenum{1}
\tablecaption{Input Physical Parameters of SN Ib/Ic Progenitor Models\label{tab:input}}
\tablehead{
\colhead{Name} & \colhead{$M_\mathrm{He,i}$} & \colhead{$M$} & 
\colhead{log $L/L_\odot$} & \colhead{$T_\star$} & \colhead{$R_\star$} &\colhead{$\dot{M}_\mathrm{fid}$} & \colhead{$v_{\infty,\mathrm{fid}}$} &
\colhead{$m_\mathrm{He}$} & \colhead{$Y$} & \colhead{log $X_\mathrm{C}$} & \colhead{log $X_\mathrm{N}$} & \colhead{log $X_\mathrm{O}$}\\ 
\colhead{} & \colhead{($M_\odot$)} & \colhead{($M_\odot$)} & \nocolhead{($L_\odot$)} & \colhead{(K)}  &  \colhead{($R_\odot$)}
& \colhead{($M_\odot \mathrm{\ yr^{-1}}$)} & \colhead{$\mathrm{(km\ s^{-1})}$} & \colhead{($M_\odot$)} 
& \colhead{} & \colhead{} & \colhead{}& \colhead{} 
}
\startdata
HE2.91 & 3.9 & 2.91 & 4.66 & 16850 & 25.02 & 2.37e-06 & 184.49 & 1.06 & 0.98 & -3.31 & -1.82 & -3.44 \\
HE2.97 & 4.0 & 2.97 & 4.68 & 19066 & 19.97 & 2.51e-06 & 206.87 & 1.07 & 0.98 & -3.31 & -1.82 & -3.44 \\
HE4.09 & 6.0 & 4.09 & 4.97 & 36312 & 7.67 & 5.49e-06 & 335.08 & 1.11 & 0.98 & -3.66 & -1.86 & -3.09 \\
HE5.05 & 8.0 & 5.05 & 5.11 & 47949 & 5.22 & 8.13e-06 & 415.33 & 0.88 & 0.98 & -3.69 & -1.86 & -3.08 \\
CO5.18 & 10.0 & 5.18 & 5.12 & 95083 & 1.34 & 1.55e-05 & 728.35 & 0.23 & 0.42 & -0.33 & - & -1.10 \\
CO5.50 & 12.0 & 5.50 & 5.18 & 117324 & 0.94 & 1.26e-05 & 942.77 & 0.17 & 0.21 & -0.27 & - & -0.65 \\
CO6.17 & 15.0 & 6.17 & 5.22 & 133296 & 0.76 & 1.41e-05 & 1157.21 & 0.20 & 0.23 & -0.27 & - & -0.68 \\
CO7.50 & 20.0 & 7.50 & 5.33 & 172779 & 0.52 & 1.70e-05 & 1715.16 & 0.18 & 0.21 & -0.28 & - & -0.62 \\
CO9.09 & 25.0 & 9.09 & 5.43 & 191741 & 0.47 & 1.95e-05 & 2171.91 & 0.16 & 0.19 & -0.29 & - & -0.56 \\
\hline
CO2.16 &   & 2.16 & 4.40 & 204316 & 0.13 & 3.25e-06 & 798.00 & 0.06 & 0.31 & -0.27 & - & -0.89 \\
CO3.93 &   & 3.93 & 4.26 & 78260 & 0.77 & 3.07e-06 & 414.08 & 0.10 & 0.49 & -0.36 & - & -1.19 \\
\enddata
\tablecomments{Input model parameters of SN Ib/Ic progenitors as provided by the stellar evolution codes (\texttt{BEC} or \texttt{MESA}). The first nine models are based on \citetalias{Yoon2017MNRAS.470.3970Y}, and the remaining two models are from \citetalias{Yoon2019ApJ...872..174Y}. From left to right, each column represents:\newline
$-M_\mathrm{He,i}$ : initial mass of the He star at the beginning of the stellar evolution code, \newline 
$-M$ : SN progenitor mass after finishing the stellar evolution code, \newline
$-L$ : total bolometric luminosity, \newline 
$-T_\star$, $R_\star$ : hydrostatic surface temperature and radius (without correcting for optical depth effects from the wind), \newline
$-\dot{M}_\mathrm{fid}$ : mass-loss rate calculated with the \citetalias{Yoon2017MNRAS.470.3970Y} prescription for the given surface properties,\newline
$-v_{\infty,\mathrm{fid}}$ : wind terminal velocity determined from the Equations (16) and (17) of \citet{Nugis2000},\newline
$-m_\mathrm{He}$ : integrated helium mass,\newline
$-Y$, $X_\mathrm{C}$, $X_\mathrm{N}$, $X_\mathrm{O}$ : surface mass fractions of helium, carbon, nitrogen and oxygen of the progenitor model.
}
\end{deluxetable*}

The physical parameters of the eleven SN Ib/Ic progenitor models used in this study are summarized in Table \ref{tab:input}. Most of the SN Ib/Ic progenitor models (the first nine models in Table \ref{tab:input}) are based on the study by \citet[][hereafter \citetalias{Yoon2017MNRAS.470.3970Y}]{Yoon2017MNRAS.470.3970Y}, in which the progenitor models are constructed using the \texttt{BEC} stellar evolution code \citep[see][and references therein]{Yoon2010ApJ...725..940Y}. These models are calculated from pure helium stars until the end of core oxygen burning at solar metallicity ($Z=0.02$). The remaining two compact progenitor models are taken from \citet[][hereafter \citetalias{Yoon2019ApJ...872..174Y}]{Yoon2019ApJ...872..174Y}, where the progenitor models are constructed from either a closed binary system (CO2.16) or a pure helium star (CO3.93) at solar metallicity using the \texttt{MESA} stellar evolution code \citep{Paxton2011ApJS..192....3P, Paxton2013ApJS..208....4P, Paxton2015ApJS..220...15P, Paxton2018ApJS..234...34P, Paxton2019ApJS..243...10P}. 

Helium-rich models with names starting with `HE' have surface helium mass fraction of $Y = 0.98$ and a substantial integrated helium mass ($m_\mathrm{HE}\equiv\int X_\mathrm{He}~dM_r \ge 0.88~M_\odot$). In contrast, helium-poor models starting with `CO' have a relatively low surface helium mass fraction ($Y < 0.5$) and a lower integrated helium mass ($m_\mathrm{He} \le 0.23~M_\odot$). We assume the `HE' and `CO' models as progenitor models of SNe Ib and Ic, respectively, following recent theoretical and observational studies indicating a dichotomy in the helium mass in SN Ib and Ic progenitors \citep{Liu2016ApJ...827...90L, Yoon2017MNRAS.470.3970Y, Yoon2019ApJ...872..174Y, Dessart2020A&A...642A.106D, Shahbandeh2022ApJ...925..175S, Jin2023ApJ...950...44J}. 

The numbers following `HE' or `CO' in the model names denote the masses of models at the pre-SN phase. The ranges of progenitor mass for SNe Ib ($2.91-5.05~M_\odot$) and Ic ($2.16-9.09~M_\odot$) are consistent with the recent observational studies on the SN Ib/Ic ejecta masses \citep[e.g.,][]{Richardson2006AJ....131.2233R, Drout2011ApJ...741...97D, Cano2013MNRAS.434.1098C, Bianco2014ApJS..213...19B, Modjaz2014AJ....147...99M, Taddia2015A&A...574A..60T, Liu2016ApJ...827...90L, Lyman2016MNRAS.457..328L, Prentice2016MNRAS.458.2973P, Taddia2018A&A...609A.136T, Prentice2019MNRAS.485.1559P, Gangopadhyay2020MNRAS.497.3770G, Barbarino2021A&A...651A..81B, Pandey2021MNRAS.507.1229P, Zheng2022MNRAS.512.3195Z}. 

Most SN Ib/Ic progenitor models (except CO2.16) used in this study evolve from a single helium star, while their progenitor mass range supports the binary-stripped progenitor scenario. The majority of SN Ib/Ic progenitors in binary systems would lose their hydrogen-rich envelopes during the so-called Case B mass transfer (i.e., mass transfer during He core contraction or early stage of core He burning). The evolution thereafter is mainly determined by stellar wind mass-loss \citep[e.g.,][]{Yoon2017ApJ...840...10Y}. Therefore, a pure helium star can approximately represent a state of SN Ib/Ic progenitors after binary stripping, with wind mass-loss playing a prominent role in their subsequent evolution. In addition, the focus of this study is to investigate the effects of winds during the pre-SN phase, rather than examining a specific evolutionary path. See \citetalias{Yoon2017MNRAS.470.3970Y} and \citetalias{Yoon2019ApJ...872..174Y} for further details on the progenitor models.

\subsection{Atmospheric Modeling} \label{sec:atmospheric}

We use the non-LTE radiative transfer code CMFGEN \citep{Hillier1998ApJ...496..407H} to generate synthetic spectra of SN Ib/Ic progenitor models. The hydrodynamic structure of the wind must be specified for the CMFGEN computation, and we assume the wind velocity profile following the standard $\beta$ law with $\beta=1$:

\begin{fleqn}[\parindent]
\begin{equation}\label{eq:betalaw}
    \begin{split}
        v(r)=v_0+(v_\infty-v_0)\left(1-\frac{R_\star}{r}\right)^\beta~~, 
    \end{split}
\end{equation}
\end{fleqn}
where $v_0$, $v_\infty$ and, $R_\star$ denote the velocity at the hydrostatic stellar surface, the terminal velocity, and the radius at the hydrostatic stellar surface, respectively \citep{Lamers1999isw..book.....L}. For the wind terminal velocities of progenitor models ($v_{\infty}$), we adopt the empirical formula for WN and WC stars from \citet{Nugis2000}. 

We determine the fiducial wind mass-loss rate ($\dot{M}_\mathrm{fid}$) of the SN Ib/Ic progenitor models using the W-R mass-loss rate prescription given by \citetalias{Yoon2017MNRAS.470.3970Y}. The prescription is derived by combining the empirical mass-loss rate for WNe-type W-R stars by the Potsdam group \citep{Hamann2006A&A...457.1015H, Hainich2014A&A...565A..27H} and for WC-type W-R stars by \citet{Tramper2016ApJ...833..133T}. Consistent with \citetalias{Yoon2017MNRAS.470.3970Y}, we adopt a wind factor $f_\mathrm{W-R}$ of 1.58 as our fiducial mass-loss rate, which better explains the observed faint WC-type galactic W-R stars.

Due to the lack of low-mass stripped star samples \citep[e.g., HD 45166;][]{Groh2008A&A...485..245G}, as we do in this study, the community has conventionally been assuming their mass-loss rates by extrapolating the W-R mass-loss prescription \citep[e.g.,][]{Yoon2012A&A...544L..11Y, Gotberg2017A&A...608A..11G, Yoon2017ApJ...840...10Y}. However, some recent theoretical studies suggest significantly lower mass-loss rates for low-mass stripped stars compared to previous extrapolated prescriptions \citep{Vink2017A&A...607L...8V,Sander2020MNRAS.491.4406S}. Moreover, the mass-loss rates of SN Ib/Ic progenitors can have significant diversity due to factors such as metallicity \citep{Vink2005A&A...442..587V} and potential mass-loss enhancements at the final stages of stellar evolution \citep{Fuller2018MNRAS.476.1853F, Sun2020MNRAS.497.5118S, Sun2020MNRAS.491.6000S, Leung2021ApJ...923...41L, Wu2021ApJ...906....3W, Sun2022MNRAS.510.3701S, Wu2022ApJ...930..119W}. Although the changes in stellar structure that could result from these phenomena are beyond the scope of this study, we account for this diversity by considering different wind mass-loss rates of $\dot{M}_\mathrm{fid}\times0.1,\;0.5,\;1.0,\;2.0,\;5.0\;\mathrm{and}\;10.0$ in our atmosphere models. The effects of uncertainties in the wind terminal velocities of SN Ib/Ic progenitors can be discussed together with the effects of diverse wind mass-loss rates, given that W-R spectra look similar for a given $R_\star(v_\infty/\dot{M})^{2/3}$ \citep{Schmutz1989A&A...210..236S}.

\section{Results of Atmospheric Models} \label{sec:results}

\subsection{Effects of winds on the SN Ib/Ic progenitor spectra}

\begin{figure*}[!htbp]
\centering
\includegraphics[width=0.95\textwidth]{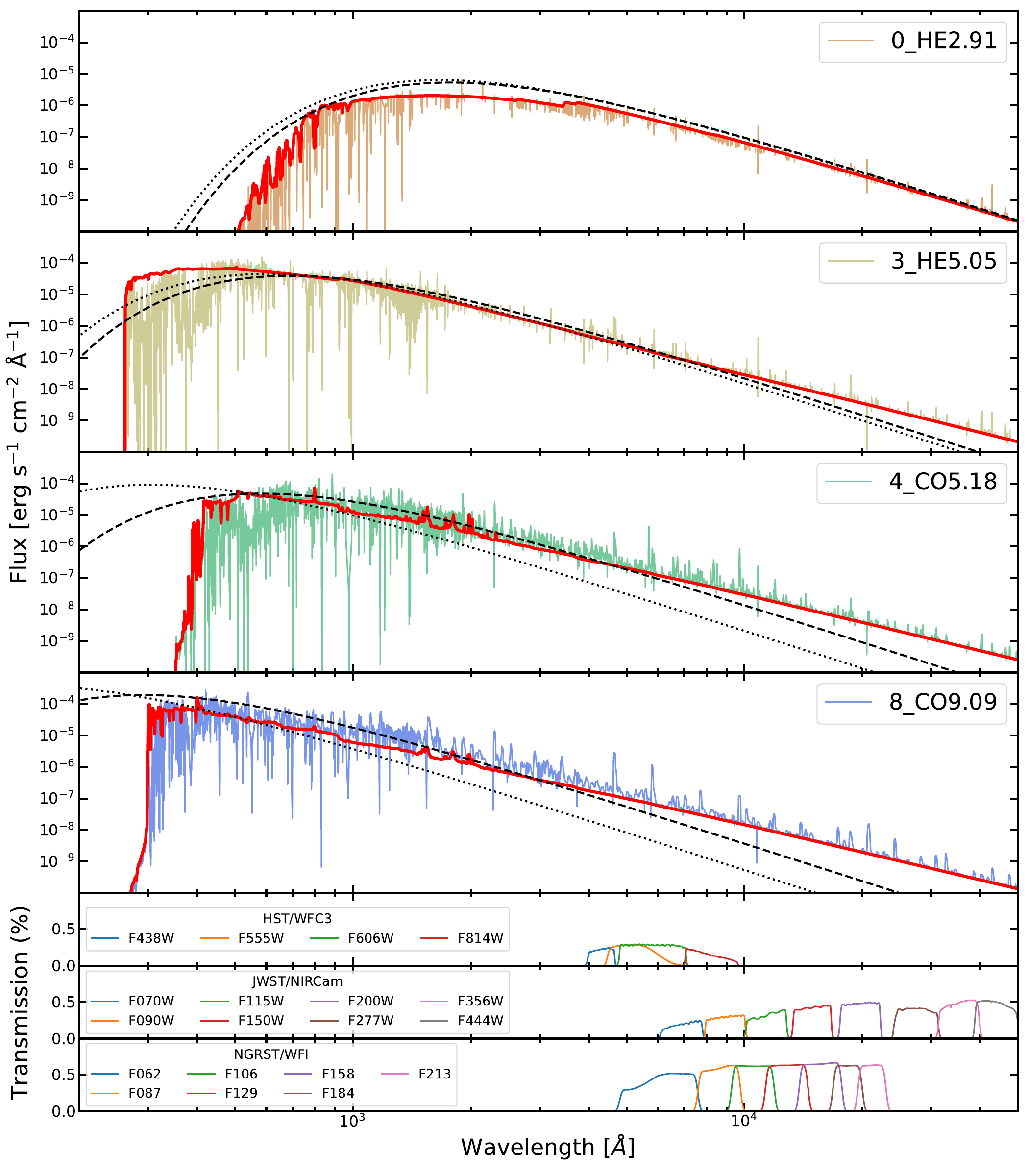}
\caption{Spectral energy distributions of four SN Ib/Ic progenitor models with the fiducial mass-loss rate (HE2.91, HE5.05, CO5.18, and CO9.09). The continuum fluxes are presented as the red line. The blackbody fluxes at the photosphere and the hydrostatic core are plotted with the dashed and dotted lines, respectively. Bottom three panels display transmission curves of broad-band filters from HST/WFC3, JWST/NIRCam, and NGRST/WFI. \label{fig:fig1_sed}}
\end{figure*}

Figure \ref{fig:fig1_sed} presents the spectral energy distributions (SEDs) of four representative SN Ib/Ic progenitor models (HE2.91, HE5.05, CO5.18, and CO9.09) with fiducial mass-loss rates. HE2.91 represents a helium giant with a stellar radius of $R_{\star,\mathrm{HE2.91}}=25.02\,R_\odot$, whereas HE5.05 is a more compact SN Ib progenitor model with $R_{\star,\mathrm{HE5.05}}=5.22\,R_\odot$. Despite their similar progenitor masses ($M_\mathrm{HE5.05}=5.05\,M_\odot$, $M_\mathrm{CO5.18}=5.18\,M_\odot$) and bolometric luminosities (log $L_\mathrm{HE5.05}/L_\odot=5.11$, log $L_\mathrm{CO5.18}/L_\odot=5.12$), HE5.05 and CO5.18 differ significantly in composition. CO5.18 is considerably more compact and hotter ($R_{\star,\mathrm{CO5.18}}=1.34\,R_\odot$, $T_{\star,\mathrm{CO5.18}}=95,083$ K) compared to HE5.05 ($R_{\star,\mathrm{HE5.05}}=5.22\,R_\odot$, $T_{\star,\mathrm{HE5.05}}=47,949$ K). CO9.09, the most massive, compact, and luminous progenitor model in this study, has a progenitor mass of $M_\mathrm{CO9.09}=9.09\,M_\odot$, a stellar radius of $R_{\star,\mathrm{CO9.09}}=0.47\,R_\odot$, and a bolometric luminosity of log $L_\mathrm{CO9.09}/L_\odot=5.43$.

In Figure \ref{fig:fig1_sed}, we present the full spectra, continuum flux, and blackbody spectra at the temperatures of the photosphere and the stellar surface to show the effects of winds. For the HE2.91 model, the differences among the full spectra, continuum flux, blackbody flux at the photosphere, and blackbody flux at the hydrostatic surface are not prominent. This indicates that the photosphere is formed close to the hydrostatic surface and the emission lines are not strong in the HE2.91 model. However, the differences are much more significant for the HE5.05 and CO models because these models are more compact and have denser winds than the HE2.91 model. First, the optically thick wind causes the photosphere to form at a radius significantly larger than the hydrostatic surface, and the corresponding effective temperature is significantly lower than the hydrostatic surface temperature. Second, the CO model spectra exhibit continuum excess at longer wavelengths compared to the blackbody flux at the photosphere. The highly ionized wind of the hot, compact models produces free-free emission, which significantly enhances the brightness at longer wavelengths, particularly in the infrared. Additionally, the compact models have numerous emission lines that can critically influence the broad-band magnitude. (See Figure \ref{fig:fig1_sed} and Chapter 3.1 in \citetalias{Jung2022ApJ...925..216J} for detailed discussions on the spectral features and effects of winds.)

\begin{splitdeluxetable*}{c|rrc|ccccBc|ccccccccBc|ccccccc}
\tablenum{2}
\tabletypesize{\scriptsize}
\tablecaption{CMFGEN Output Parameters and Optical/Near-IR Magnitudes of SN Ib/Ic Progenitor Models \label{tab:ABmag}}
\tablehead{
\multicolumn{1}{c|}{}
 & \multicolumn{3}{c|}{} & \multicolumn{4}{c}{HST/WFC3}
 & \multicolumn{1}{c|}{}  & \multicolumn{8}{c}{JWST/NIRCam} 
 & \multicolumn{1}{c|}{}  & \multicolumn{7}{c}{NGRST/WFI} \\
\multicolumn{1}{c|}{Model} 
 & \colhead{$T_\mathrm{eff}$} & \colhead{$R_\mathrm{phot}$} & \multicolumn{1}{c|}{log $L/L_\odot$}& \colhead{F438W} & \colhead{F555W} & \colhead{F606W} & \colhead{F814W} 
 & \multicolumn{1}{c|}{Model}  & \colhead{F070W} & \colhead{F090W} & \colhead{F115W} & \colhead{F150W} & \colhead{F200W} & \colhead{F277W} & \colhead{F356W} & \colhead{F444W} 
 & \multicolumn{1}{c|}{Model}  & \colhead{F062} & \colhead{F087} & \colhead{F106} & \colhead{F129} & \colhead{F158} & \colhead{F184} & \colhead{F213} \\
\multicolumn{1}{c|}{} 
& \colhead{(K)} & \colhead{($R_\star$)} & \multicolumn{1}{c|}{} 
& \colhead{(ABmag)} & \colhead{(ABmag)} & \colhead{(ABmag)} & \colhead{(ABmag)}
& \multicolumn{1}{c|}{}  & \colhead{(ABmag)} & \colhead{(ABmag)} & \colhead{(ABmag)} & \colhead{(ABmag)} & \colhead{(ABmag)} & \colhead{(ABmag)} & \colhead{(ABmag)} & \colhead{(ABmag)}
& \multicolumn{1}{c|}{}  & \colhead{(ABmag)} & \colhead{(ABmag)} & \colhead{(ABmag)} & \colhead{(ABmag)} & \colhead{(ABmag)} & \colhead{(ABmag)} & \colhead{(ABmag)}
}
\startdata
HE2.91 & 16330 & 26.65 & 4.66 & -5.39 & -5.22 & -5.11 & -4.74 & HE2.91 & -4.89 & -4.60 & -4.25 & -3.83 & -3.37 & -2.78 & -2.32 & -2.04 & HE2.91 & -5.04 & -4.64 & -4.39 & -4.07 & -3.75 & -3.49 & -3.24 \\ 
HE2.97 & 18390 & 21.46 & 4.68 & -5.25 & -5.04 & -4.93 & -4.50 & HE2.97 & -4.67 & -4.35 & -3.99 & -3.54 & -3.09 & -2.52 & -2.08 & -1.85 & HE2.97 & -4.85 & -4.39 & -4.13 & -3.80 & -3.47 & -3.22 & -2.97 \\ 
HE4.09 & 34730 & 8.38 & 4.97 & -4.90 & -4.59 & -4.41 & -3.90 & HE4.09 & -4.12 & -3.70 & -3.42 & -2.92 & -2.60 & -2.11 & -1.75 & -1.71 & HE4.09 & -4.32 & -3.76 & -3.56 & -3.17 & -2.86 & -2.73 & -2.45 \\ 
HE5.05 & 41910 & 6.83 & 5.11 & -4.74 & -4.45 & -4.26 & -3.88 & HE5.05 & -4.05 & -3.71 & -3.58 & -3.14 & -2.92 & -2.49 & -2.17 & -2.13 & HE5.05 & -4.19 & -3.75 & -3.68 & -3.33 & -3.10 & -3.03 & -2.77 \\ 
CO5.18 & 49850 & 4.88 & 5.12 & -5.04 & -4.85 & -4.67 & -4.31 & CO5.18 & -4.44 & -4.39 & -3.79 & -3.32 & -3.15 & -2.70 & -2.34 & -2.30 & CO5.18 & -4.62 & -4.35 & -4.18 & -3.59 & -3.29 & -3.25 & -2.97 \\ 
CO5.50 & 68200 & 2.78 & 5.18 & -4.72 & -4.51 & -4.15 & -3.83 & CO5.50 & -3.97 & -4.07 & -3.34 & -3.05 & -2.88 & -2.38 & -2.09 & -1.95 & CO5.50 & -4.09 & -3.99 & -3.90 & -3.24 & -3.05 & -2.93 & -2.77 \\ 
CO6.17 & 73450 & 2.51 & 5.22 & -4.76 & -4.56 & -4.13 & -3.79 & CO6.17 & -3.95 & -3.97 & -3.35 & -3.04 & -2.93 & -2.40 & -2.11 & -1.97 & CO6.17 & -4.07 & -3.93 & -3.86 & -3.24 & -3.05 & -2.95 & -2.85 \\ 
CO7.50 & 89770 & 1.91 & 5.33 & -4.66 & -4.61 & -4.14 & -3.69 & CO7.50 & -3.86 & -3.60 & -3.25 & -2.90 & -2.92 & -2.30 & -1.99 & -1.88 & CO7.50 & -4.07 & -3.68 & -3.55 & -3.13 & -2.92 & -2.87 & -2.88 \\ 
CO9.09 & 99710 & 1.73 & 5.43 & -4.54 & -4.48 & -3.99 & -3.54 & CO9.09 & -3.72 & -3.47 & -3.08 & -2.71 & -2.76 & -2.11 & -1.80 & -1.68 & CO9.09 & -3.92 & -3.55 & -3.41 & -2.96 & -2.74 & -2.71 & -2.72 \\ 
\hline
CO2.16 & 74890 & 0.94 & 4.40 & -2.84 & -2.63 & -2.31 & -2.02 & CO2.16 & -2.13 & -2.26 & -1.52 & -1.20 & -0.99 & -0.54 & -0.23 & -0.08 & CO2.16 & -2.25 & -2.19 & -2.07 & -1.40 & -1.19 & -1.06 & -0.89 \\ 
CO3.93 & 48080 & 1.95 & 4.26 & -3.04 & -2.82 & -2.68 & -2.35 & CO3.93 & -2.50 & -2.34 & -1.85 & -1.36 & -1.21 & -0.77 & -0.41 & -0.40 & CO3.93 & -2.64 & -2.33 & -2.16 & -1.63 & -1.33 & -1.32 & -1.02 \\
\enddata
\tablecomments{The model names correspond to those listed in Table \ref{tab:input}. $T_\mathrm{eff}$ represents the effective temperature, and $R_\mathrm{phot}$ denotes the photosphere radius defined at the Rosseland optical depth $\tau=2/3$. The remaining columns show the AB magnitudes of SN Ib/Ic progenitor models with fiducial mass-loss rates, as measured through broad-band filters on the HST/WFC3, JWST/NIRCam, and NGRST/WFI.}
\end{splitdeluxetable*}

\citetalias{Jung2022ApJ...925..216J} showed the optical data of SN Ib/Ic progenitors and compared them with the SN Ib/Ic progenitor search results from the HST and other optical ground-based telescopes. In this study, we shift our focus to the effects of winds in the near-IR and investigate the possibilities and advantages of near-IR observations. The James Webb Space Telescope (JWST) is providing deep near-IR images with higher spatial resolution than the HST, and the Nancy Grace Roman Space Telescope (NGRST), expected to be launched in 2026-2027, will provide vast data with a field of view $\sim$100 times wider ($\sim0.28\,\mathrm{deg}^2$) than that of the HST ($\sim0.002\,\mathrm{deg}^2$ for UVIS channel of Wide Field Camera3 (WFC3)) or JWST ($\sim0.003\,\mathrm{deg}^2$ for Near Infrared Camera (NIRCam)). 

In Table \ref{tab:ABmag}, we present the magnitudes of SN Ib/Ic progenitor models with fiducial mass-loss rates in the broad-band filters of the JWST/NIRCam and NGRST Wide-Field Instrument (WFI). The HST/WFC3 magnitudes are also given in the table for comparison. The CMFGEN output parameters, including the effective temperature ($T_\mathrm{eff}$) and the photosphere radius ($R_\mathrm{phot}$) defined at the Rosseland optical depth of $\tau=2/3$, are presented in the same table. The filter magnitudes for HST, JWST and NGRST are computed using \texttt{synphot} and \texttt{stsynphot} Python packages. Compared to optical filter magnitudes, infrared magnitudes in the $\sim2\mu\mathrm{m}$ and $\sim4\mu\mathrm{m}$ filters increase by about $1-2$ and $2-3$ mag, respectively. Notably, the absolute magnitude difference between the HE2.91 and CO9.09 models is about 1 mag in the optical filters, but reduces to only 0.36 mag in the JWST/F444W filter. This underscores the significant impact of stellar winds on SN Ib/Ic progenitor spectra, as previously discussed.

\subsection{Effects of the mass-loss rates}

\begin{figure*}[!htbp]
\includegraphics[width=0.95\textwidth]{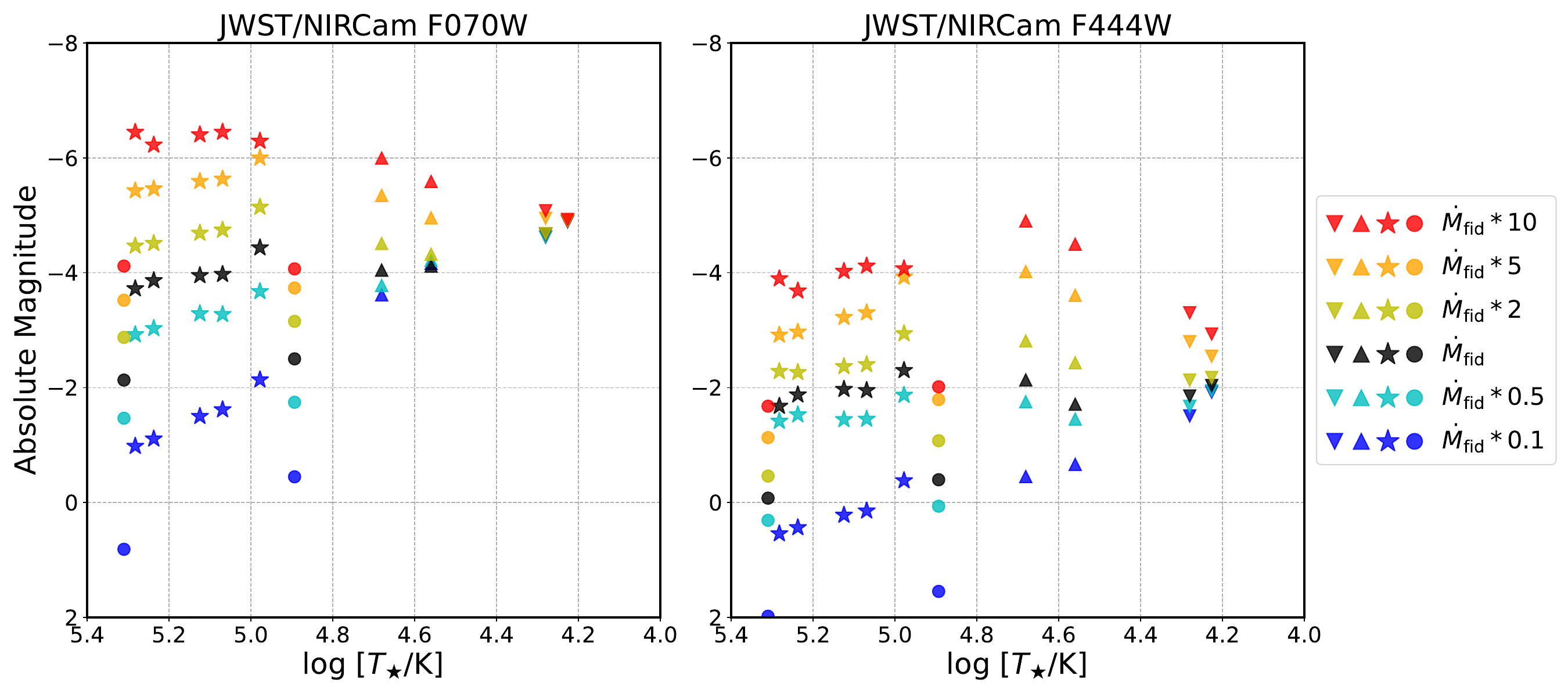}
\centering
\caption{Absolute magnitudes of SN Ib/Ic progenitors for various mass-loss rates. The upper and lower panels show absolute magnitudes for the JWST/NIRCam F070W and F444W filters, respectively. Inverted triangles represent giant HE models, while upright triangles depict compact HE models, both from \citetalias{Yoon2017MNRAS.470.3970Y}. CO models are presented as star symbols for those from \citetalias{Yoon2017MNRAS.470.3970Y} and filled circles for those from \citetalias{Yoon2019ApJ...872..174Y}. Each color of the markers correspond to the adopted mass-loss rate as specified in the legend. \label{fig:fig2_Mdot}}
\end{figure*}

As mentioned earlier, we considered various mass-loss rates ($0.1-10$ times of $\dot{M}_\mathrm{fid}$) to account for the highly uncertain mass-loss rates of SN Ib/Ic progenitors. The absolute magnitudes of SN Ib/Ic progenitor models with different mass-loss rates in the JWST F070W and F444W filters are presented in Figure \ref{fig:fig2_Mdot}. (The absolute magnitudes of these progenitor models with various mass-loss rates in HST/WFC3, JWST/NIRCam, and NGRST/WFI filters are summarized in Appendix \ref{sec:appendixA}.) This result demonstrate that the models with higher mass-loss rates are significantly brighter in the near-IR, consistent with the optical results (see Chapter 3.2 of \citetalias{Jung2022ApJ...925..216J}). For SN Ic progenitors, a mass-loss rate difference by 100 times results in a magnitude difference of $\sim5$, independent of wavelength. 

For SN Ib progenitors, the magnitude differences resulting from changes in the mass-loss rate increase more significantly in the longer wavelength filters compared to SN Ic progenitors. Giant HE models (HE2.91 and HE2.97), denoted by inverted triangles in Figure \ref{fig:fig2_Mdot}, show similar JWST/F070W filter magnitudes across different mass-loss rates, but the effects of varying mass-loss rates become prominent in the JWST/F444W filter. Even for compact HE models (HE4.09 and HE5.05), denoted by upright triangles, the magnitude differences increase from $\sim2-3$  magnitudes in F070W to over 5 magnitudes in F444W for the considered $\dot{M}$ range. These results suggest that near-IR magnitudes can be used to constrain the mass-loss rate for SN Ib/Ic progenitors including helium giant stars.

\section{Detectability of SN Ib/Ic progenitors with JWST and NGRST} \label{sec:detectability}

\subsection{Comparison of HST, JWST, and NGRST} \label{subsec:HST,JWST,NGRST}

Over the last 30 years, the HST has directly detected dozens of SN II progenitors, but only one progenitor and two candidates for SNe Ib/Ic have been identified. In our previous study of \citetalias{Jung2022ApJ...925..216J}, we argue that the three directly identified SN Ib/Ic progenitors or progenitor candidates should have high mass-loss rates ($\dot{M} > 5\times\dot{M}_\mathrm{fid}$) or a bright companion, and the exposure time of $t_\mathrm{exp}<20$ min is too short to detect progenitors for the fiducial mass-loss rate. Fortunately, JWST, which is a powerful near-IR telescope, is now available for the community, and another near-IR space telescope, NGRST, is expected to be launched within a few years. Observations of hot stars have advantages in the optical over near-IR observations as they are much brighter in short wavelengths. However, as discussed in Section \ref{sec:results} above, SN Ib/Ic progenitors would be fairly bright in near-IR if the mass-loss rate was sufficiently high, so we discuss the detection possibility with the three space telescopes here. 

\begin{figure}[!htbp]
\includegraphics[width=0.48\textwidth]{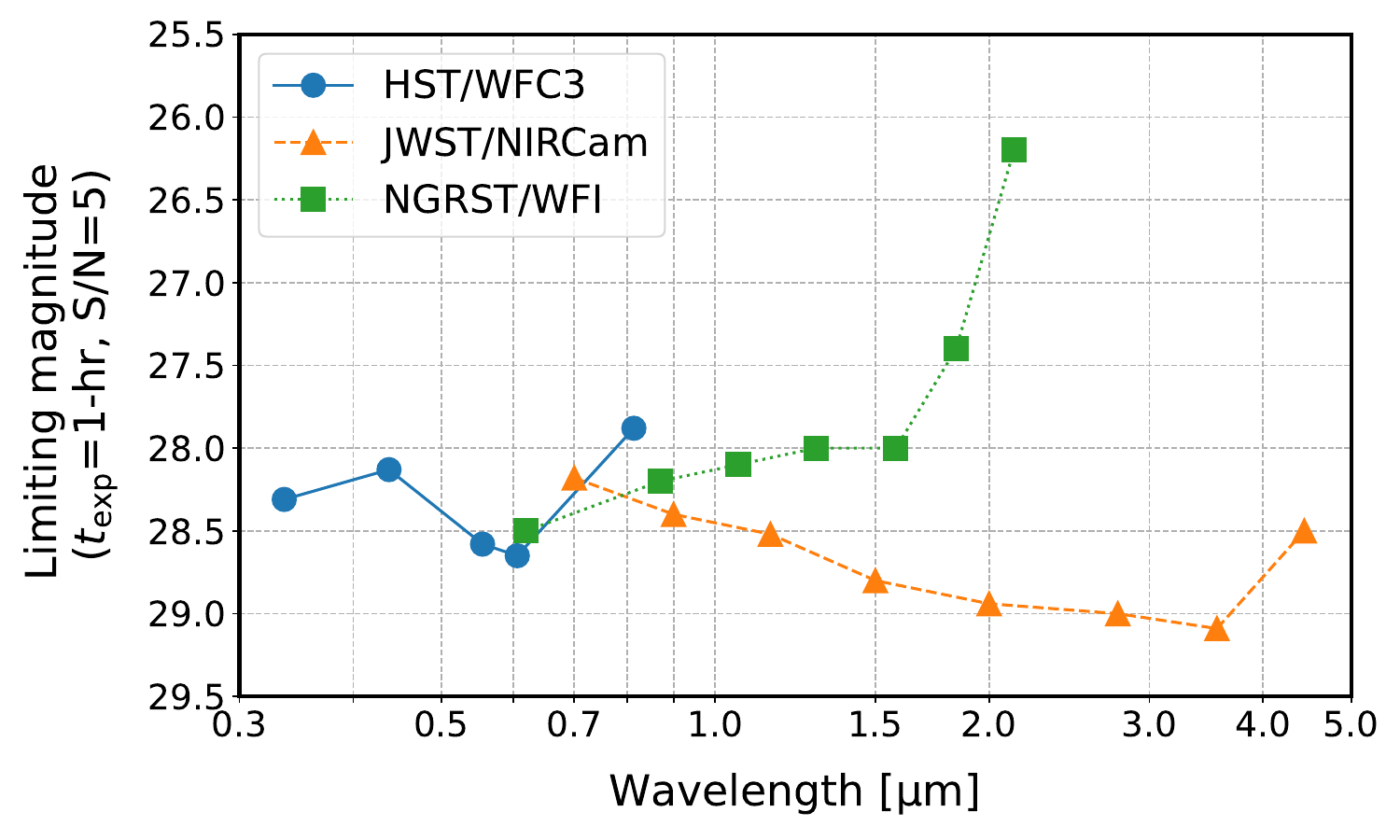}
\centering
\caption{Limiting magnitudes of broad-band filters in HST/WFC3, JWST/NIRCam and NGRST/WFI required to achieve a signal-to-noise ratio (S/N) of 5 with a 1 hour exposure. \label{fig:fig3_limit}}
\end{figure}

\begin{figure*}[!htbp]
\includegraphics[width=0.9\textwidth]{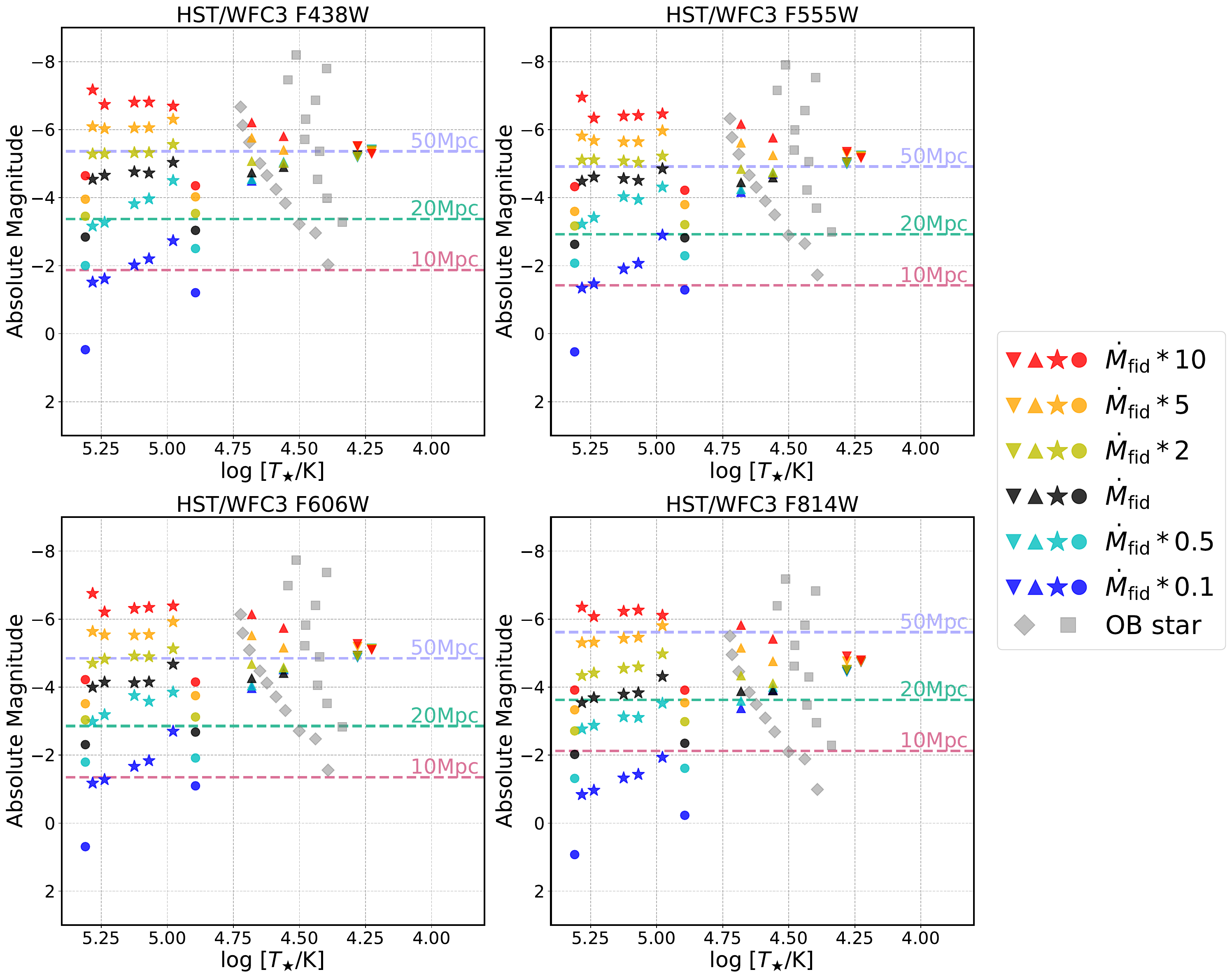}
\centering
\caption{HST/WFC3 broad-band filter magnitudes of SN Ib/Ic progenitors with various mass-loss rates. Extinction is not applied to the models presented in this figure. Stars and triangles denote the HE and CO progenitor models from \citetalias{Yoon2017MNRAS.470.3970Y}, respectively, while the CO progenitor models from \citetalias{Yoon2019ApJ...872..174Y} are presented with circles. For comparison, optical magnitudes of O-type and B-type (O/B-type) stars with initial masses ($M_\mathrm{init}$) of 9, 12, 15, 20, 25, 32, 40, 60, 85, and 120 $M_\odot$ are presented together in gray diamonds (ZAMS models) and squares (evolved models) for the comparison. The O/B-type star models are taken from \citet{Fierro2015PASP..127..428F}. \label{fig:fig4_HST_detect}}
\end{figure*}

The limiting magnitudes for various broad-band filters of HST/WFC3, JWST/NIRCam and NGRST/WFI are presented in Figure \ref{fig:fig3_limit}. The JWST has comparable limiting magnitudes compared to the HST in filters with $\mathrm{\lambda_{center}< 1.3\mu m}$, but surpasses the HST by approximately 1 mag at longer wavelengths. The limiting magnitudes of the NGRST are comparable to the HST in filters with $\mathrm{\lambda_{center} < 1.8\mu m}$ but are less sensitive at longer wavelengths. The limiting magnitudes are calculated to achieve a signal-to-noise ratio (S/N) of 5 with a 1 hour exposure time for a point source with a flat continuum in $F_{\lambda}$ under standard background conditions. Extinction has not been considered in these calculations. The limiting magnitudes for HST/WFC3 filters are calculated using the HST Exposure Time Calculator (ver. 32.2)\footnote{\url{https://etc.stsci.edu/etc/input/wfc3uvis/imaging/}}. The values for JWST/NIRCam filters are obtained using the JWST Exposure Time Calculator (ver. 3.2)\footnote{\url{https://jwst.etc.stsci.edu/}} with the \texttt{DEEP8} readout pattern, 4 groups per integration, and 5 integrations per exposure. For NGRST/WFI filters, their limiting magnitudes are taken from the anticipated performance tables\footnote{\url{https://roman.gsfc.nasa.gov/science/apttables2021/table-exposuretimes.html}} for a point source under a Zodiacal Light background level of 1.2 times the minimum.

\begin{figure*}[!htbp]
\includegraphics[width=0.7\textwidth]{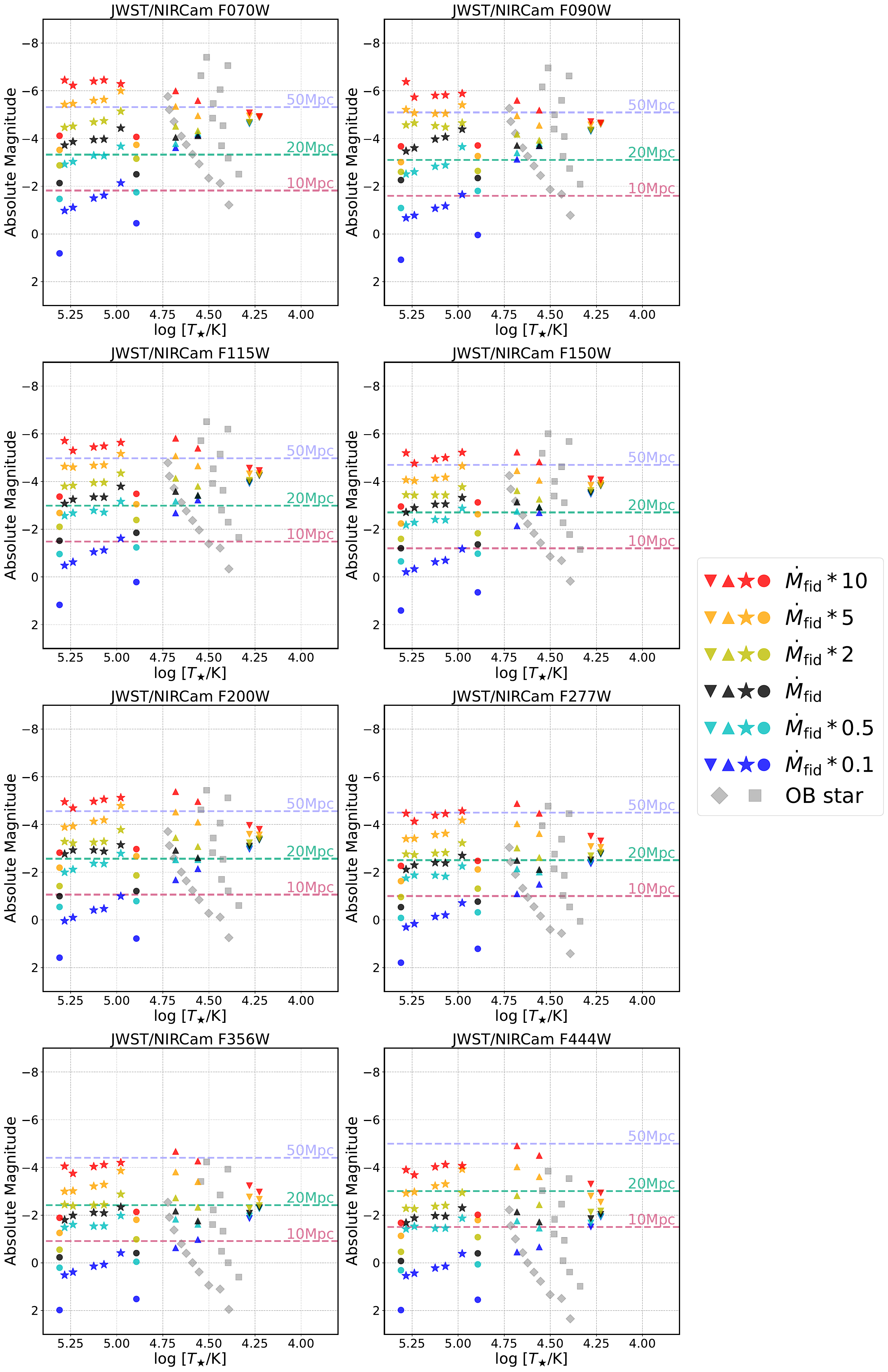}
\centering
\caption{JWST/NIRCam broad-band filter magnitudes of SN Ib/Ic progenitors with various mass-loss rates without considering the effects of extinction. The markers and colors have the same meanings as those in Figure \ref{fig:fig4_HST_detect}. 
 \label{fig:fig5_JWST_detect}}
\end{figure*}

\begin{figure*}[!htbp]
\includegraphics[width=0.7\textwidth]{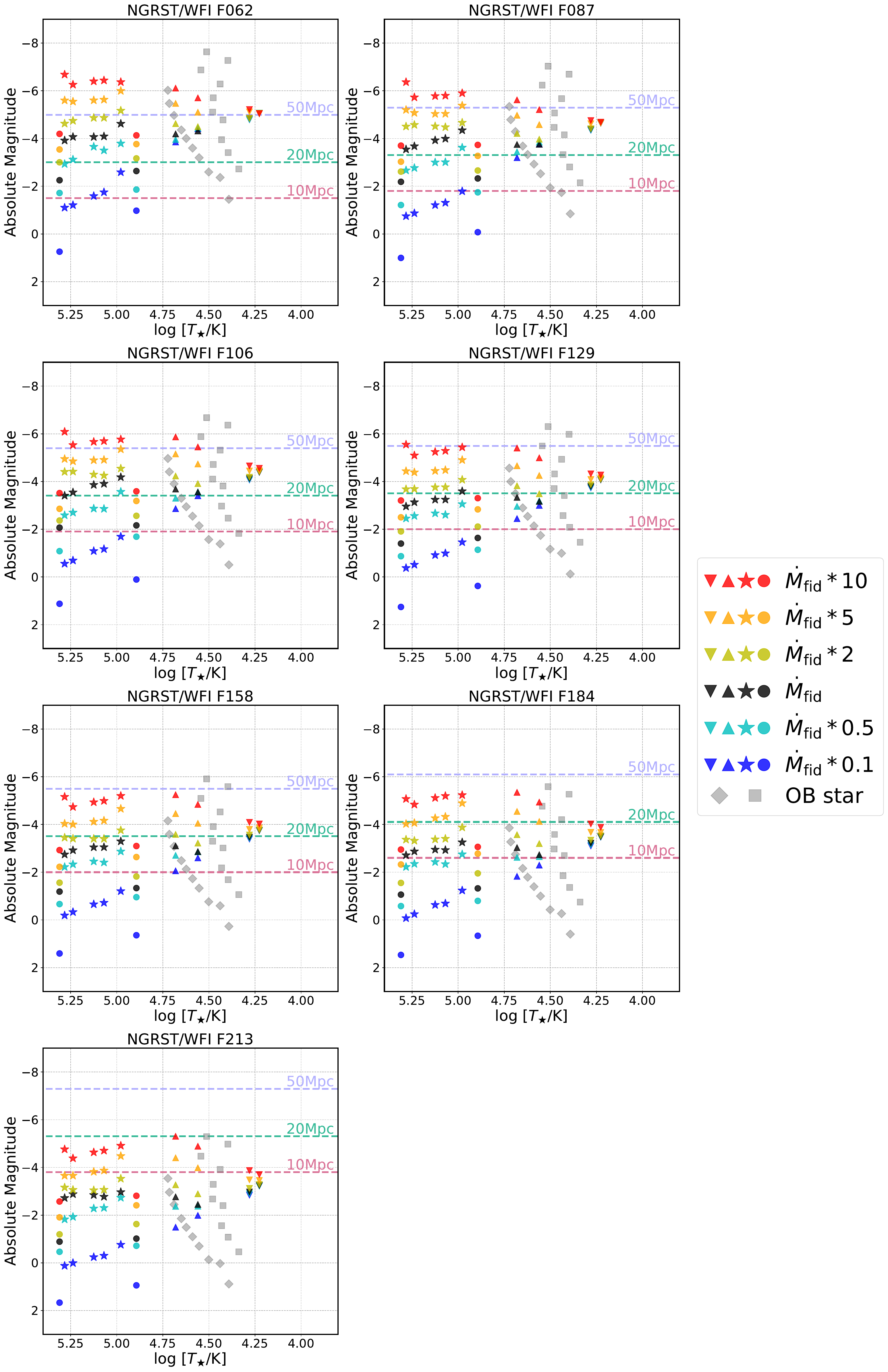}
\centering
\caption{NGRST/WFI broad-band filter magnitudes of SN Ib/Ic progenitors with various mass-loss rates without considering the effects of extinction. The markers and colors have the same meanings as those in Figure \ref{fig:fig4_HST_detect}.  \label{fig:fig6_NGRST_detect}}
\end{figure*}

\begin{deluxetable*}{ccccc}
\tablenum{3}
\tabletypesize{\scriptsize}
\tablecaption{Constraints on mass-loss rates for detectable SN Ib/Ic progenitors at 20 Mpc ($t_\mathrm{exp}=$ 1 hour; $A_\mathrm{V}=0$ and 1 mag) \label{tab:limit}}
\tablehead{}
\startdata
\hline
\multicolumn{5}{c}{$A_\mathrm{V}=0$ mag} \\ \hline
\multicolumn{1}{c|}{Filters} & \multicolumn{1}{c}{\begin{tabular}[c]{@{}c@{}}SN Ic prog\\ (Y19)\end{tabular}} & \begin{tabular}[c]{@{}c@{}}SN Ic prog\\ (Y17)\end{tabular} & \begin{tabular}[c]{@{}c@{}}Compact SN Ib prog\\ (Y17)\end{tabular} & \begin{tabular}[c]{@{}c@{}}Giant SN Ib prog\\ (Y17)\end{tabular} \\ \hline
\multicolumn{1}{l|}{\begin{tabular}[c]{@{}l@{}}F555W (HST)\\ F606W (HST)\end{tabular}} & $>2\times\dot{M}_\mathrm{fid}$ & $>0.5\times\dot{M}_\mathrm{fid}$ & $>0.1\times\dot{M}_\mathrm{fid}$ & $>0.1\times\dot{M}_\mathrm{fid}$ \\ \hline
\multicolumn{1}{l|}{\begin{tabular}[c]{@{}l@{}}F438W (HST)\end{tabular}} & $>2\times\dot{M}_\mathrm{fid}$ & $>\dot{M}_\mathrm{fid}$ & $>0.1\times\dot{M}_\mathrm{fid}$ & $>0.1\times\dot{M}_\mathrm{fid}$ \\ \hline
\multicolumn{1}{l|}{\begin{tabular}[c]{@{}l@{}}F070W (JWST)\\ F062 (NGRST)\end{tabular}} & $>5\times\dot{M}_\mathrm{fid}$ & $>\dot{M}_\mathrm{fid}$ & $>0.1\times\dot{M}_\mathrm{fid}$ & $>0.1\times\dot{M}_\mathrm{fid}$ \\ \hline
\multicolumn{1}{l|}{\begin{tabular}[c]{@{}l@{}}F090W (JWST)\\F115W (JWST)\\F150W (JWST)\\F087 (NGRST) \end{tabular}} & $>10\times\dot{M}_\mathrm{fid}$ & $>\dot{M}_\mathrm{fid}$ & $>0.5\times\dot{M}_\mathrm{fid}$ & $>0.1\times\dot{M}_\mathrm{fid}$ \\ \hline
\multicolumn{1}{l|}{\begin{tabular}[c]{@{}l@{}}F200W (JWST)\\F106 (NGRST) \end{tabular}} & $>10\times\dot{M}_\mathrm{fid}$ & $>\dot{M}_\mathrm{fid}$ & $>\dot{M}_\mathrm{fid}$ & $>0.1\times\dot{M}_\mathrm{fid}$ \\ \hline
\multicolumn{1}{l|}{F814W (HST)} & $>10\times\dot{M}_\mathrm{fid}$ & $>2\times\dot{M}_\mathrm{fid}$ & $>\dot{M}_\mathrm{fid}$ & $>0.1\times\dot{M}_\mathrm{fid}$ \\  \hline
\multicolumn{1}{l|}{F277W (JWST)} & & $>2\times\dot{M}_\mathrm{fid}$ & $>2\times\dot{M}_\mathrm{fid}$ & $>\dot{M}_\mathrm{fid}$ \\ 
\multicolumn{1}{l|}{F129 (NGRST)} & & $>2\times\dot{M}_\mathrm{fid}$ & $>5\times\dot{M}_\mathrm{fid}$ & $>0.1\times\dot{M}_\mathrm{fid}$ \\ \hline
\multicolumn{1}{l|}{F158 (NGRST)} & & $>5\times\dot{M}_\mathrm{fid}$ & $>5\times\dot{M}_\mathrm{fid}$ & $>2\times\dot{M}_\mathrm{fid}$ \\ \hline
\multicolumn{1}{l|}{F356W (JWST)} & & $>5\times\dot{M}_\mathrm{fid}$ & $>5\times\dot{M}_\mathrm{fid}$ & $>5\times\dot{M}_\mathrm{fid}$ \\ \hline
\multicolumn{1}{l|}{\begin{tabular}[c]{@{}l@{}}F444W (JWST) \end{tabular}} & & $>10\times\dot{M}_\mathrm{fid}$ & $>5\times\dot{M}_\mathrm{fid}$ & \\ \hline
\multicolumn{1}{l|}{\begin{tabular}[c]{@{}l@{}}F184 (NGRST) \end{tabular}} & & $>10\times\dot{M}_\mathrm{fid}$ & $>10\times\dot{M}_\mathrm{fid}$ & \\ \hline
\multicolumn{1}{l|}{F213 (NGRST)} & & & & \\ \hline
& & & & \\ \hline
\multicolumn{5}{c}{$A_\mathrm{V}=1$ mag} \\ \hline
\multicolumn{1}{c|}{Filters} & \multicolumn{1}{c}{\begin{tabular}[c]{@{}c@{}}SN Ic prog\\ (Y19)\end{tabular}} & \begin{tabular}[c]{@{}c@{}}SN Ic prog\\ (Y17)\end{tabular} & \begin{tabular}[c]{@{}c@{}}Compact SN Ib prog\\ (Y17)\end{tabular} & \begin{tabular}[c]{@{}c@{}}Giant SN Ib prog\\ (Y17)\end{tabular} \\ \hline
\multicolumn{1}{l|}{\begin{tabular}[c]{@{}l@{}}F200W (JWST)\end{tabular}} & & $>2\times\dot{M}_\mathrm{fid}$ & $>2\times\dot{M}_\mathrm{fid}$ & $>0.1\times\dot{M}_\mathrm{fid}$ \\ \hline
\multicolumn{1}{l|}{\begin{tabular}[c]{@{}l@{}}F150W (JWST)\end{tabular}} & & $>2\times\dot{M}_\mathrm{fid}$ & $>5\times\dot{M}_\mathrm{fid}$ & $>0.1\times\dot{M}_\mathrm{fid}$ \\ \hline
\multicolumn{1}{l|}{\begin{tabular}[c]{@{}l@{}}F115W (JWST)\end{tabular}} & & $>5\times\dot{M}_\mathrm{fid}$ & $>5\times\dot{M}_\mathrm{fid}$ & $>2\times\dot{M}_\mathrm{fid}$ \\ \hline
\multicolumn{1}{l|}{\begin{tabular}[c]{@{}l@{}}F277W (JWST)\end{tabular}} & & $>5\times\dot{M}_\mathrm{fid}$ & $>5\times\dot{M}_\mathrm{fid}$ & $>5\times\dot{M}_\mathrm{fid}$ \\ \hline
\multicolumn{1}{l|}{\begin{tabular}[c]{@{}l@{}}F356W (JWST)\\F106 (NGRST)\end{tabular}} & & $>5\times\dot{M}_\mathrm{fid}$ & $>5\times\dot{M}_\mathrm{fid}$ & $>10\times\dot{M}_\mathrm{fid}$ \\ \hline
\multicolumn{1}{l|}{\begin{tabular}[c]{@{}l@{}}F090W (JWST)\end{tabular}} & & $>5\times\dot{M}_\mathrm{fid}$ & $>10\times\dot{M}_\mathrm{fid}$ & $>10\times\dot{M}_\mathrm{fid}$ \\ \hline
\multicolumn{1}{l|}{\begin{tabular}[c]{@{}l@{}}F087 (NGRST)\end{tabular}} & & $>5\times\dot{M}_\mathrm{fid}$ & $>10\times\dot{M}_\mathrm{fid}$ & \\
\multicolumn{1}{l|}{\begin{tabular}[c]{@{}l@{}}F129 (NGRST)\end{tabular}} & & $>5\times\dot{M}_\mathrm{fid}$ & $>10\times\dot{M}_\mathrm{fid}$ & \\
\multicolumn{1}{l|}{\begin{tabular}[c]{@{}l@{}}F444W (JWST)\end{tabular}} & & $>10\times\dot{M}_\mathrm{fid}$ & $>5\times\dot{M}_\mathrm{fid}$ & \\ \hline
\multicolumn{1}{l|}{\begin{tabular}[c]{@{}l@{}}F062 (NGRST)\\F158 (NGRST)\\F184 (NGRST)\\F213 (NGRST)\end{tabular}} & & $>10\times\dot{M}_\mathrm{fid}$ & $>10\times\dot{M}_\mathrm{fid}$ & \\ \hline
\multicolumn{1}{l|}{\begin{tabular}[c]{@{}l@{}}F555W (HST)\\F606W (HST)\\F814W (HST) \\ F070W (JWST)\end{tabular}} & & $>10\times\dot{M}_\mathrm{fid}$ & & \\ \hline
\multicolumn{1}{l|}{F438W (HST)} & & & & \\ \hline
\hline
\enddata
\tablecomments{The mass-loss rate ranges for detectable SN Ib/Ic progenitors using broad-band filters in the HST/WFC3, JWST/NIRCam, and NGRST/WFI. The assumed progenitor distance is 20 Mpc, and the exposure time is set to 1 hour. Empty cells in the table indicate that even models with mass-loss rate 10 times the fiducial value are fainter than the detection limit.}
\end{deluxetable*}

In Figure \ref{fig:fig4_HST_detect}, the absolute magnitudes of SN Ib/Ic progenitor models and detection limits are presented for HST/WFC3 filters. Assuming a distance of 20 Mpc, all SN Ib progenitor models are predicted to be brighter than the detection limit in the F438W, F555W, and F606W filters in HST/WFC3, but some low mass-loss rate SN Ib progenitor models (HE4.09 and HE5.05 with $\dot{M}<0.5\times\dot{M}_\mathrm{fid}$) are fainter than the detection limit in the F814W filter. Similarly, SN Ic progenitor models from \citetalias{Yoon2017MNRAS.470.3970Y} with $\dot{M}>0.5\times\dot{M}_\mathrm{fid}$ are brighter than the detection limit in the F555W and F606W filters in HST/WFC3, but the mass-loss rate limit increases to $\dot{M}>\dot{M}_\mathrm{fid}$ for the F438 and F814W filters. (Note that SN Ic progenitor models from \citetalias{Yoon2019ApJ...872..174Y} have significantly lower bolometric luminosity than the models from \citetalias{Yoon2017MNRAS.470.3970Y}, so the required mass-loss rate to be brighter than the detection limit is substantially higher for \citetalias{Yoon2019ApJ...872..174Y} models.) From Figure \ref{fig:fig4_HST_detect}, we can conclude that the SN Ib/Ic progenitors with the fiducial mass-loss rate can be detected in all HST/WFC3 filters, although detectability with the F814W filter is slightly lower than the other three filters.

Figures \ref{fig:fig5_JWST_detect} and \ref{fig:fig6_NGRST_detect} are the same as Figure \ref{fig:fig4_HST_detect}, but for filters in JWST/NIRCam and NGRST/WFI. In Figure \ref{fig:fig5_JWST_detect}, the detectability with the JWST/NIRCam F070W, F090W, F115W, F150W, and F200W filters is slightly better than that with the HST/WFC3 F814W filter, but it is worse than the detectability achieved with the other HST/WFC3 filters. With JWST/NIRCam filters with $\lambda_\mathrm{center} \geq \mathrm{2\mu m}$, the detectability decreases significantly due to the low flux of SN Ib/Ic progenitors at longer wavelengths.

As shown in Figure \ref{fig:fig6_NGRST_detect}, the detectability for SN Ib/Ic progenitors with NGRST/WFI F062 filter is comparable to that with the HST/WFC3 F438W and JWST/NIRCam F070W  filters, with which SN Ib/Ic progenitors with $\dot{M}>\dot{M}_\mathrm{fid}$ at a distance of 20 Mpc can be detected. Detectability with NGRST/WFI F087 and F106 filters are slightly better than that with the HST/WFC3 F814W filter, while using NGRST/WFI filters with $\lambda_\mathrm{center} \geq \mathrm{1.2\mu m}$ is less effective than shorter wavelength filters for SN Ib/Ic progenitor detection due to their reduced flux at longer wavelengths. The lower limits of mass-loss rates for detectable SN Ib/Ic progenitors with the HST/WFC3, JWST/NIRCam, and NGRST/WFI filters are summarized in the upper part of Table \ref{tab:limit}.

\subsection{Effect of Extinction on near-IR Detection}

\begin{figure}[!htbp]
\includegraphics[width=0.48\textwidth]{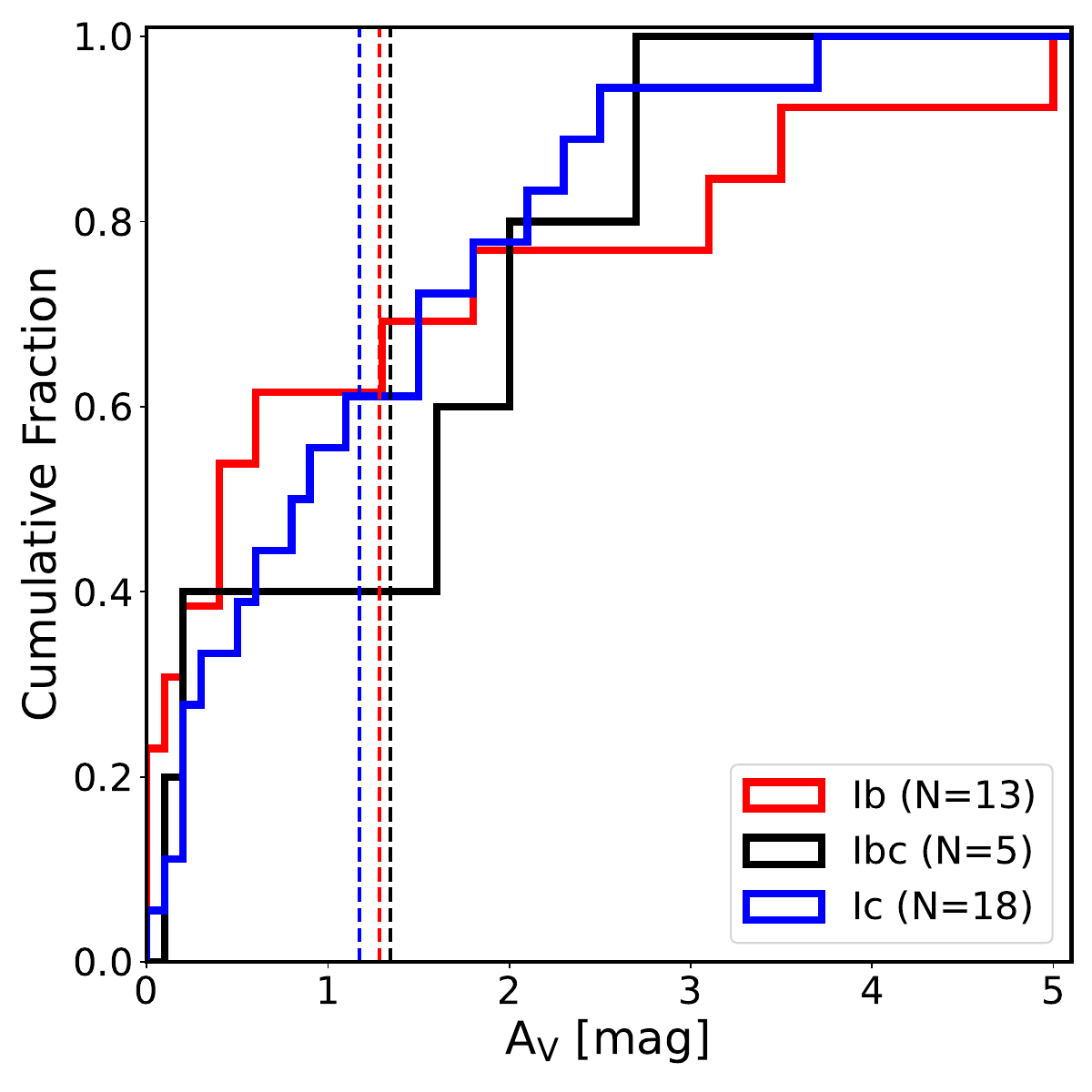}
\centering
\caption{The cumulative distribution of $A_\mathrm{V}$ for SN Ib/Ic host galaxies. The vertical dashed lines indicate the mean $A_\mathrm{V}$ value for each supernova type. The extinction data are taken from \citet{Sanders2012ApJ...758..132S} and \citet{Eldridge2013MNRAS.436..774E}. \label{fig:fig7_Av_distribution}}
\end{figure}

The results of Section \ref{subsec:HST,JWST,NGRST} seem to indicate that JWST and NGRST might not be as useful as HST in the search for SN Ib/Ic progenitors. However, it is necessary to consider the effect of extinction, which is ignored in the previous discussion. To address this, we collected V-band extinction ($A_\mathrm{V}$) values of SN Ib/Ic host galaxies from \citet{Sanders2012ApJ...758..132S} and \citet{Eldridge2013MNRAS.436..774E}, and their cumulative distribution is shown in Figure \ref{fig:fig7_Av_distribution}. The mean $A_\mathrm{V}$ of SN Ib/Ic host galaxies is significantly high, approximately $1.2-1.3$ mag, and $\sim30\%$ of them have extinctions greater than $A_\mathrm{V}=2$ mag. This implies that extinction may play an important role in SN Ib/Ic progenitor searches, and near-IR observations can be competitive in this regard.

\begin{figure*}[!htbp]
\includegraphics[width=0.95\textwidth]{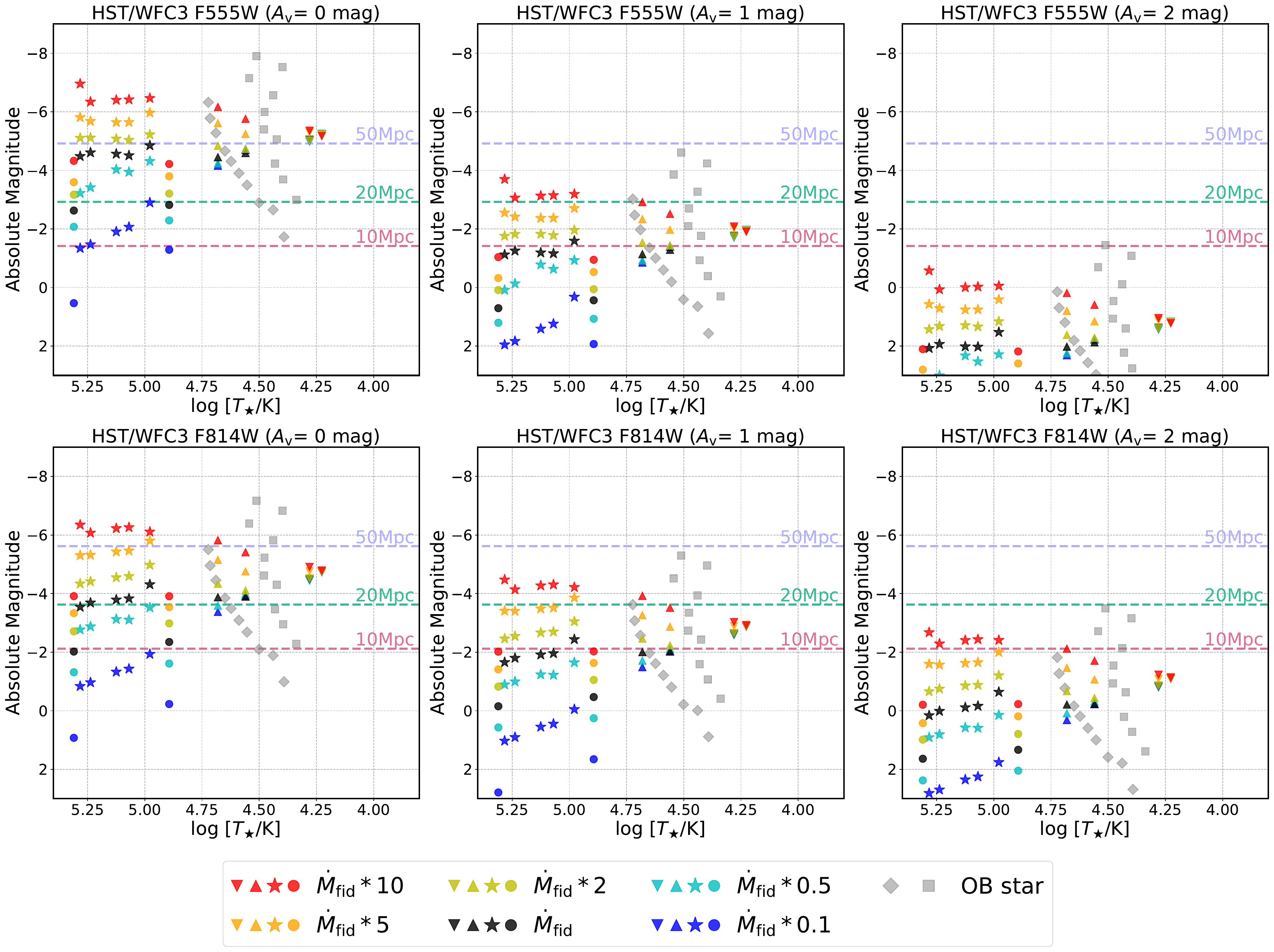}
\centering
\caption{Absolute magnitudes of SN Ib/Ic progenitor models in selected HST/WFC3, JWST/NIRCam, and NGRST/WFI filters considering extinction effects. The first, second, and third columns show the magnitudes of SN Ib/Ic progenitors under extinctions of $A_\mathrm{V}=0$, 1, and 2 mag, respectively. The meanings of each marker and color are the same as in Figure \ref{fig:fig4_HST_detect}. \label{fig:fig8_Av_effect}}
\end{figure*}

\begin{figure*}[!htbp]
\addtocounter{figure}{-1}
\renewcommand{\thefigure}{\arabic{figure} (continued)}
\includegraphics[width=0.95\textwidth]{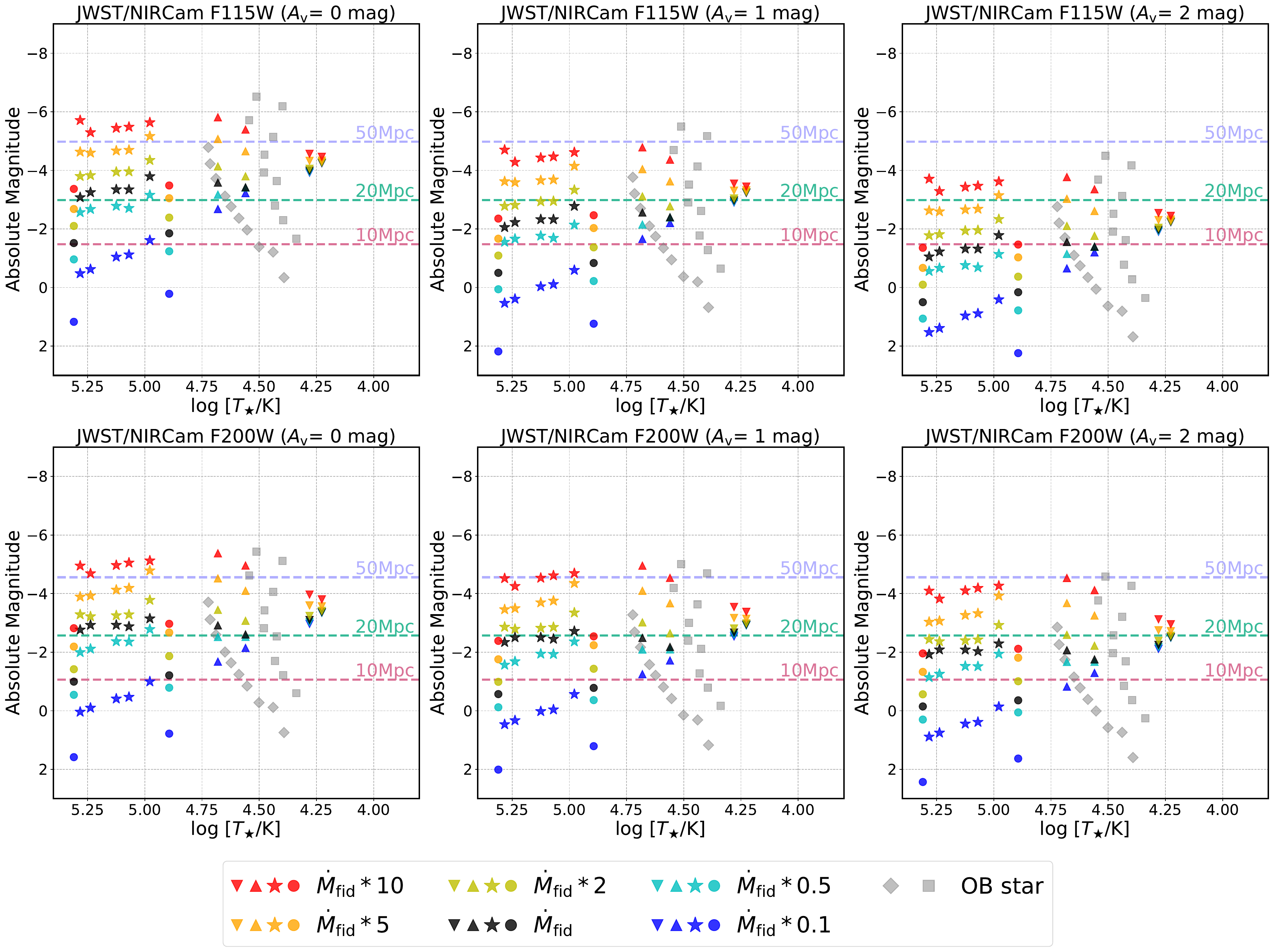}
\centering
\caption{Absolute magnitudes of SN Ib/Ic progenitor models in selected HST/WFC3, JWST/NIRCam, and NGRST/WFI filters considering extinction effects. \label{fig:fig8_2_Av_effect}}
\end{figure*}

\begin{figure*}[!htbp]
\addtocounter{figure}{-1}
\renewcommand{\thefigure}{\arabic{figure} (continued)}
\includegraphics[width=0.95\textwidth]{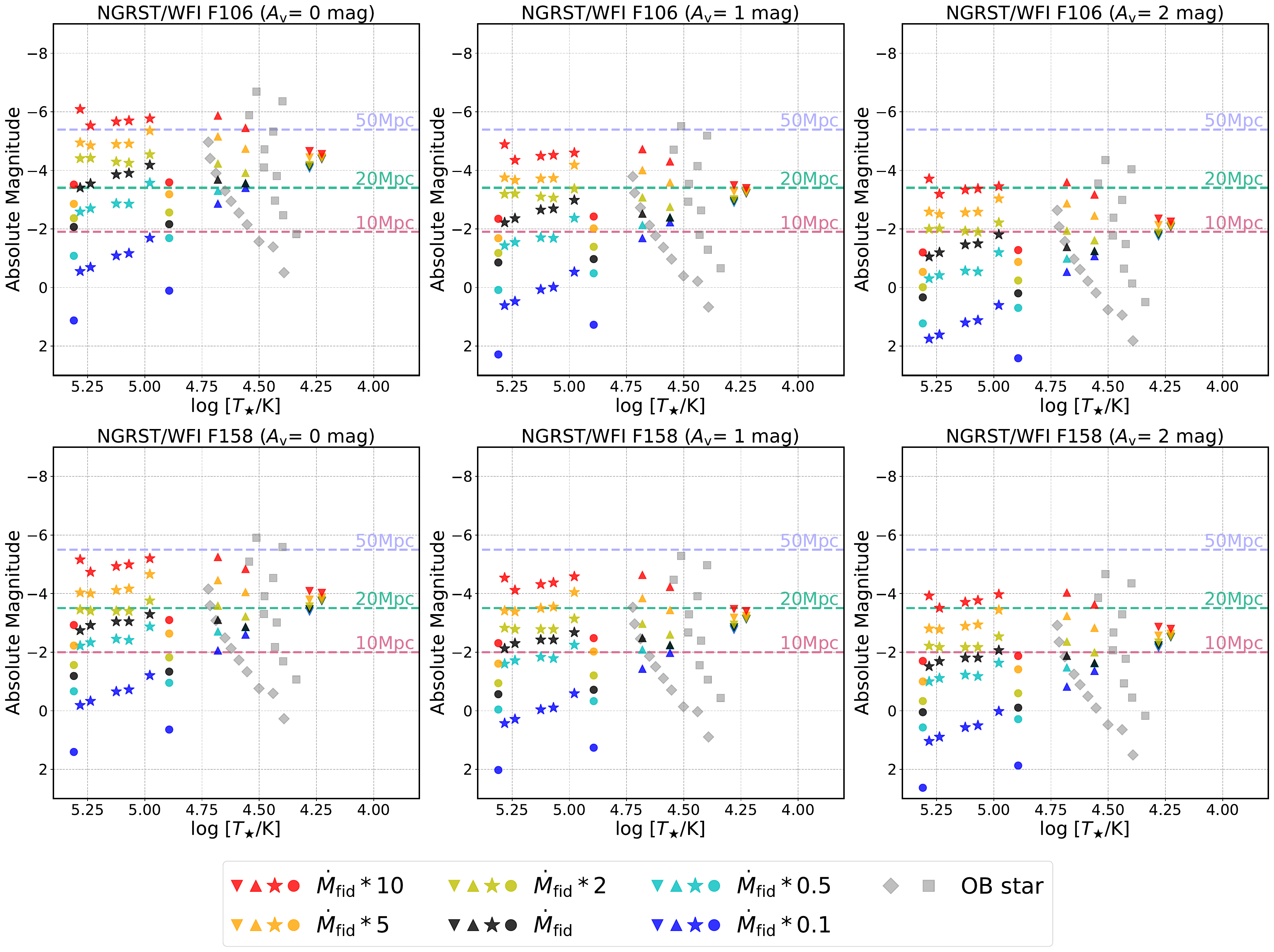}
\centering
\caption{Absolute magnitudes of SN Ib/Ic progenitor models in selected HST/WFC3, JWST/NIRCam, and NGRST/WFI filters considering extinction effects. \label{fig:fig8_3_Av_effect}}
\end{figure*}

The significant impact of extinction on the brightness of SN Ib/Ic progenitors is shown in Figure \ref{fig:fig8_Av_effect}, using six filters: F555W and F814W in HST/WFC3, F115W and F200W in JWST/NIRCam, and F106 and F158 in NGRST/WFI. For the HST F555W filter and $A_\mathrm{V}=1$ mag, only SN Ic progenitors with mass-loss rates exceeding $10\times\dot{M}_\mathrm{fid}$ are detectable at a distance of $d=20$ Mpc. For $A_\mathrm{V}=2$ mag, using the HST F555W filter, not a single SN Ib/Ic progenitor model is brighter than the detection limit, even at $d=10$ Mpc. The case is similar for the HST F814W filter, although SN Ic progenitors with mass-loss rates of $10\times\dot{M}_\mathrm{fid}$ at $d=10$ Mpc can be detected using the HST F814W filter with an extinction of $A_\mathrm{V}=2$ mag. 

The impact of extinction is significantly less pronounced for near-IR filters in JWST and NGRST. For the JWST F115W and NGRST F106 filters, $5\times\dot{M}_\mathrm{fid}$ progenitor models at $d=20$ Mpc with $A_\mathrm{V}=1$ mag are brighter than their detection limits, and some fiducial mass-loss rate models at $d=10$ Mpc with $A_\mathrm{V}=2$ mag still have magnitudes comparable to the detection limits. Since the extinction at $\sim2\mu$m is approximately half of that in the V band, some SN Ib/Ic progenitors with the fiducial mass-loss rate at $d=20$ Mpc and $A_\mathrm{V}=1$ mag can be detected using the JWST F200W filter. As shown in Figure \ref{fig:fig7_Av_distribution}, an obscured environment with $A_\mathrm{V}>1$ mag is common for SNe Ib/Ic, and we would have a higher chance to detect a SN Ib/Ic progenitor with JWST or NGRST than with HST. The range of mass-loss rates for detectable SN Ib/Ic progenitors under $A_\mathrm{V}=1$ mag is presented in the lower panel of Table \ref{tab:limit}. For $A_\mathrm{V}=1$ mag, JWST has the highest detectability for SN Ib/Ic progenitors, followed by NGRST, whereas HST cannot detect SN Ib progenitors with mass-loss rates of $10\times\dot{M}_\mathrm{fid}$.

Figure \ref{fig:fig8_Av_effect} and Table \ref{tab:limit} also provide the information about the exposure time required for SN Ib/Ic progenitor detection with HST, JWST and NGRST. Assuming $A_\mathrm{V}=1$ mag, some SN Ib/Ic progenitors at $d=10$ Mpc with a mass-loss rate comparable to $\dot{M}_\mathrm{fid}$ can be detected through optical observations with $t_\mathrm{exp}$ of 1 hour. HST observations with $t_\mathrm{exp}>$ 1 hour for nearby galaxies within 10 Mpc could be an effective strategy for future SN Ib/Ic progenitor searches, although the number of SN Ib/Ic in such close galaxies is expected to be small. With JWST and NGRST, less observation time is needed to cover the galaxies at $d<10$ Mpc, or more distant galaxies can be observed for the same exposure time compared to HST.

\section{Advantages of near-IR observation for SN Ib/Ic progenitor study} \label{sec:advantage}

\subsection{Constraint on the mass-loss rate of SN Ib/Ic progenitors}

\begin{figure*}[!htbp]
\includegraphics[width=0.95\textwidth]{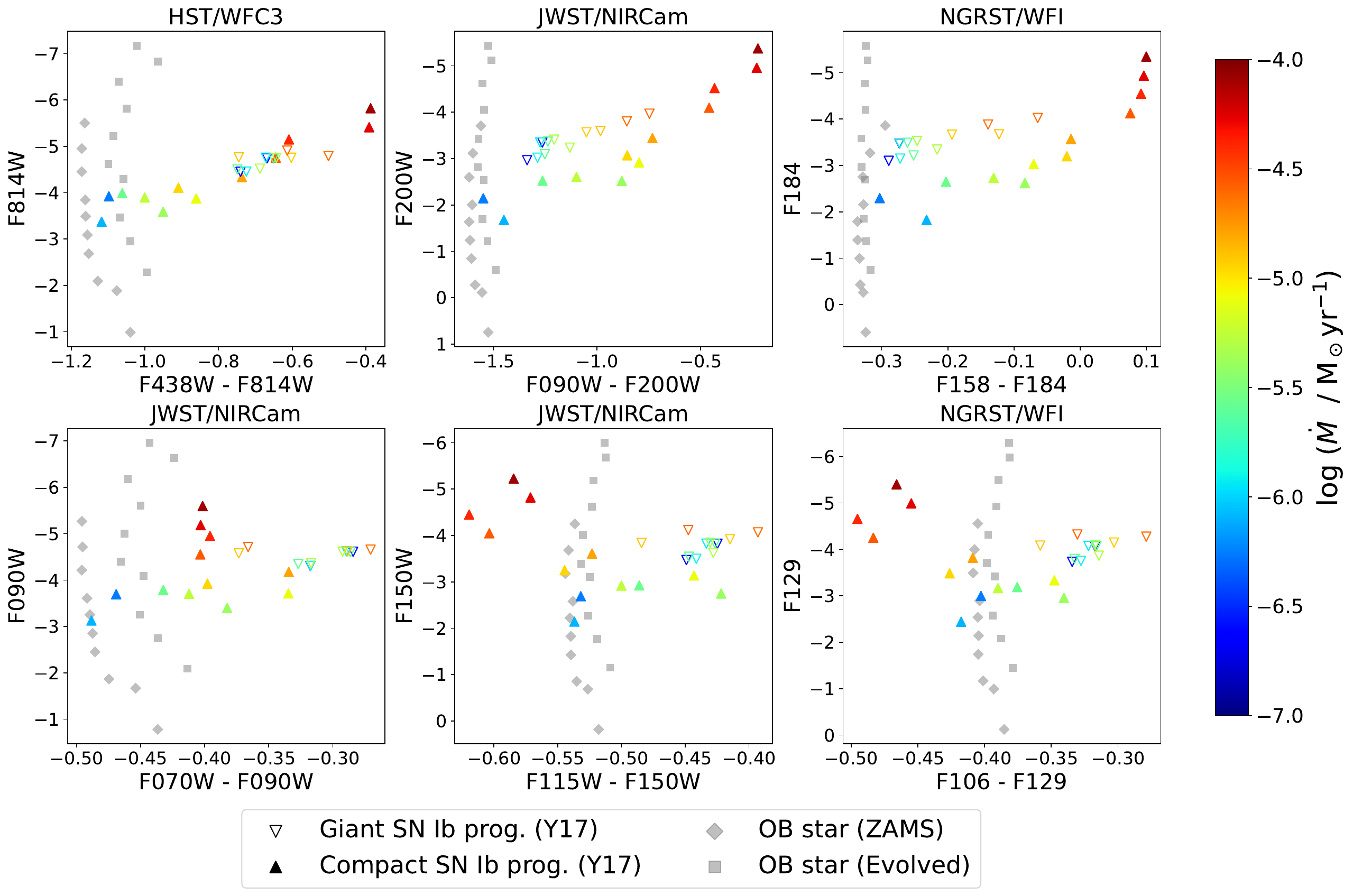}
\centering
\caption{Color-magnitude diagram of SN Ib progenitor models with various mass-loss rates. Inverted open triangles and upright filled triangles represent the giant and compact SN Ib progenitor models, respectively. The marker colors indicate the mass-loss rate. Gray diamonds and squares denote ZAMS and evolved O/B-type stars, respectively. The O/B-type star models from \citet{Fierro2015PASP..127..428F} have initial masses of 9, 12, 15, 20, 25, 32, 40, 60, 85, and 120 $M_\odot$. \label{fig:fig9_CMD_Ib}}
\end{figure*}

\begin{figure*}[!htbp]
\includegraphics[width=0.95\textwidth]{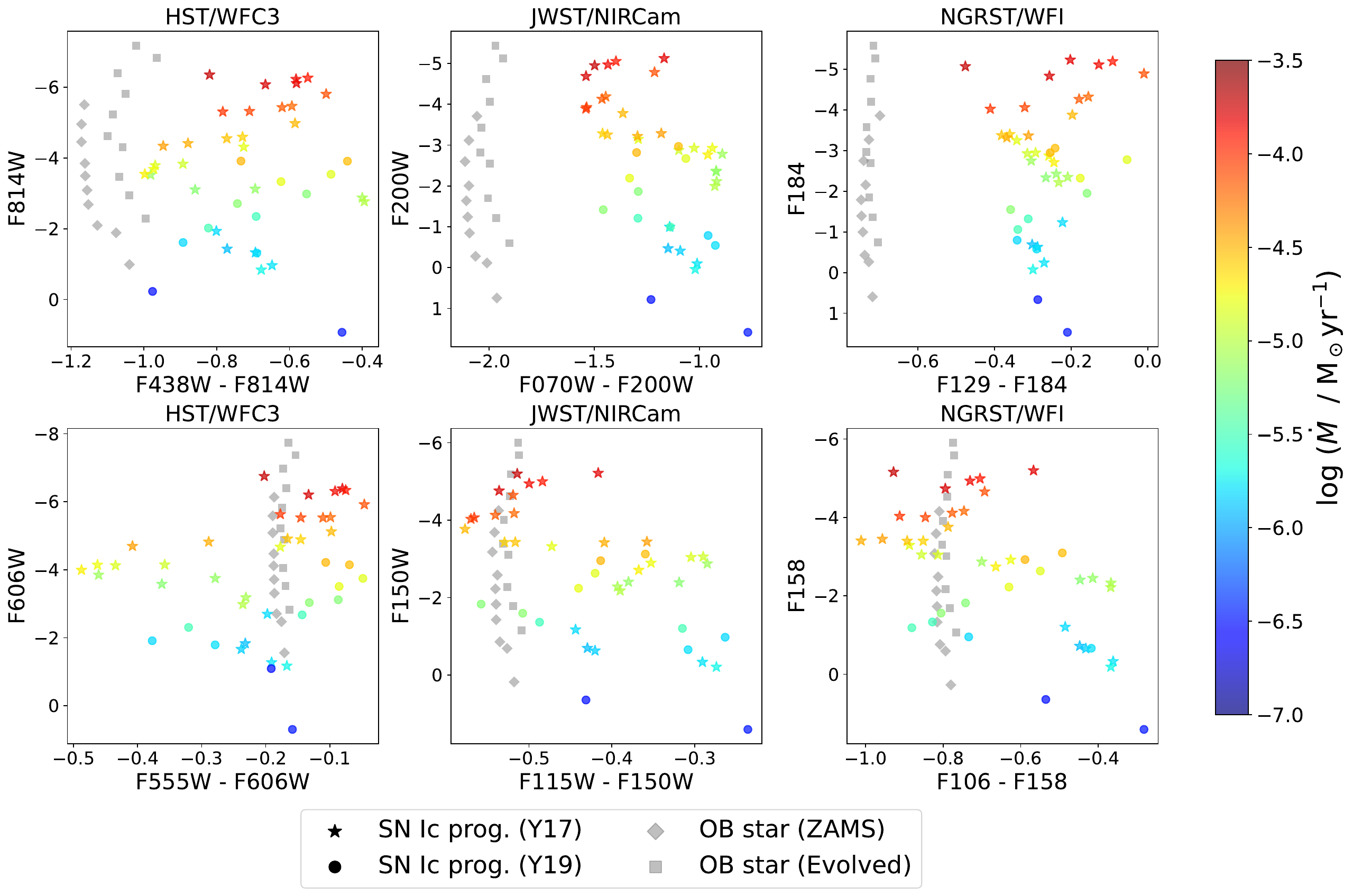}
\centering
\caption{Color-magnitude diagram of SN Ic progenitor models with various mass-loss rates. Stars and circles represent the SN Ic progenitor models from \citetalias{Yoon2017MNRAS.470.3970Y} and \citetalias{Yoon2019ApJ...872..174Y}, respectively. The marker colors indicate the mass-loss rate. Gray diamonds and squares denote ZAMS and evolved O/B-type stars, respectively. The O/B-type star models from \citet{Fierro2015PASP..127..428F} have initial masses of 9, 12, 15, 20, 25, 32, 40, 60, 85, and 120 $M_\odot$. \label{fig:fig10_CMD_Ic}}
\end{figure*}

The study by \citetalias{Jung2022ApJ...925..216J} demonstrates that the mass-loss rates of SN Ic and compact SN Ib progenitors can be well constrained using the optical filter magnitudes, as shown in Figure \ref{fig:fig4_HST_detect}. However, the variations in the optical magnitudes of giant SN Ib progenitors (HE2.91 and HE2.97 with log $(T_\star/\mathrm{K})<4.3)$ exhibit low sensitivity to changes of the mass-loss rate, making it difficult to constrain their mass-loss rates. 
 
The challenge of constraining the mass-loss rate of giant SN Ib progenitors can be overcome through near-IR observations, as denser winds result in stronger free-free emission, thereby increasing near-IR brightness. For JWST, the magnitude differences between giant SN Ib progenitors with 0.1 and $10\times\dot{M}_\mathrm{fid}$ are less than 1 magnitude in the F070W filter but increase to about 2 magnitudes in F200W and longer wavelength filters, as shown in Figure \ref{fig:fig5_JWST_detect}. The advantage of the near-IR observation is further demonstrated for the NGRST results, as shown in Figure \ref{fig:fig6_NGRST_detect}.
 
Selected color-magnitude diagrams of SN Ib progenitors are presented in Figure \ref{fig:fig9_CMD_Ib}. As shown in the upper three panels of Figure \ref{fig:fig9_CMD_Ib}, SN Ib progenitor models with higher mass-loss rates have lower magnitudes and redder colors across various optical and near-IR filters, suggesting that not only magnitudes but also colors can be used to constrain the mass-loss rates of SN Ib progenitors. However, strong emission lines can occasionally distort the color-$\dot{M}$ trend as seen in the lower three panels of Figure \ref{fig:fig9_CMD_Ib}. SN Ib progenitors generally have redder color than O/B-type stars due to free-free emission, but strong emission lines in specific filters can make the progenitor colors even bluer than those of O/B-type stars. Therefore, the colors of SN Ib progenitors should be used with caution. While compact and giant SN Ib progenitors have a distinct color-$\dot{M}$ trends, they are only distinguishable in the selected near-IR color-magnitude diagrams, as shown in the upper middle and upper right panels of Figure \ref{fig:fig9_CMD_Ib}. In contrast, compact and giant SN Ib progenitors overlap and are difficult to distinguish in the HST color-magnitude diagrams, as shown in the upper left panel of Figure \ref{fig:fig9_CMD_Ib}.

Color-magnitude diagrams of SN Ic progenitors are shown in Figure \ref{fig:fig10_CMD_Ic}. Mass-loss rates of SN Ic progenitor models can be effectively constrained by both optical and near-IR magnitudes since their effective temperature and photospheric radius are highly sensitive to the mass-loss rate. However, emission lines of SN Ic progenitors are significantly stronger than those of SN Ib progenitors, resulting in a weak or no correlation between color and mass-loss rates for SN Ic progenitors.

\subsection{Distinction between SN Ib/Ic progenitors and companion/background stars}

\begin{figure*}[!htbp]
\includegraphics[width=0.95\textwidth]{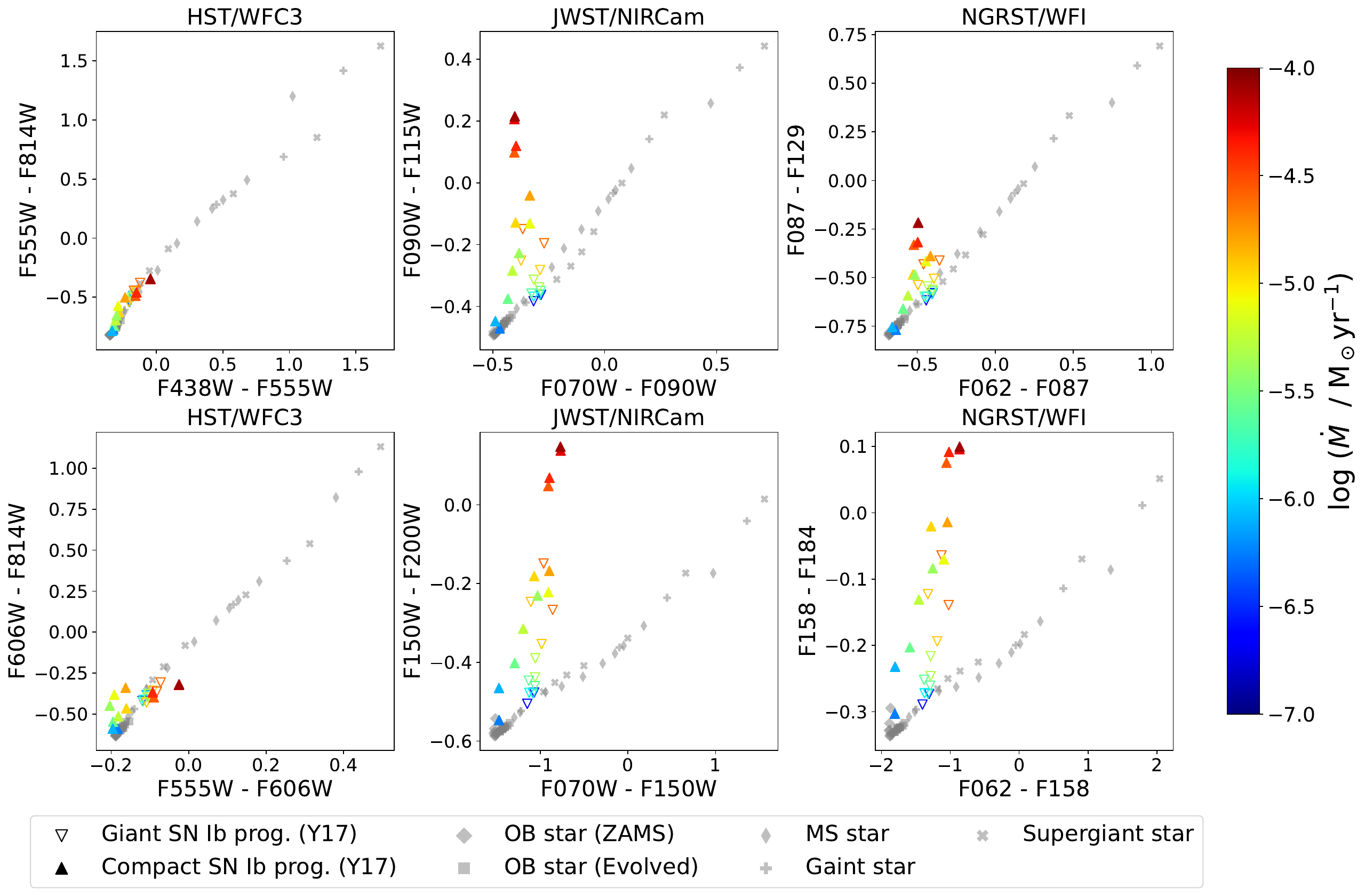}
\centering
\caption{Color-color diagrams of SN Ib progenitor models with various mass-loss rates. Inverted open triangles and upright filled triangles represent the giant and compact SN Ib progenitor models, respectively. The marker colors indicate the mass-loss rate. Diamonds and squares represent O/B-type stars in the ZAMS and evolve phase \citep{Fierro2015PASP..127..428F}. Thin diamond, plus, and cross markers denote ZAMS, giant, and supergiant stars from ATLAS9 models by \citet{Castelli2003IAUS..210P.A20C}. \label{fig:fig11_CCD_Ib}}
\end{figure*}

\begin{figure*}[!htbp]
\includegraphics[width=0.95\textwidth]{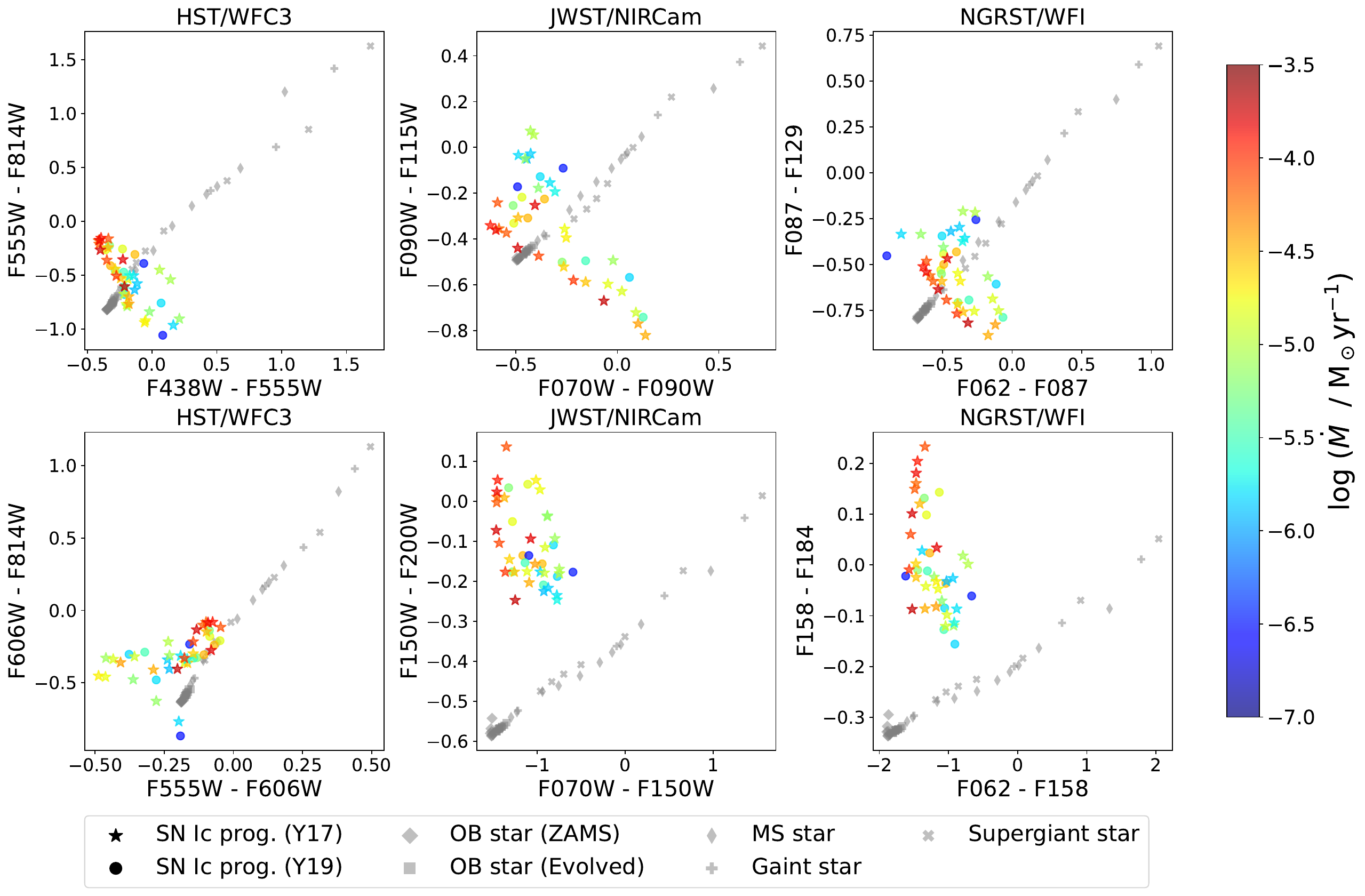}
\centering
\caption{Color-color diagrams of SN Ic progenitor models with various mass-loss rates. Stars and circles represent the SN Ic progenitor models from \citetalias{Yoon2017MNRAS.470.3970Y} and \citetalias{Yoon2019ApJ...872..174Y}, respectively. Diamonds and squares represent O/B-type stars in the ZAMS and evolve phase \citep{Fierro2015PASP..127..428F}. Thin diamond, plus, and cross markers denote ZAMS, giant, and supergiant stars, respectively, based on ATLAS9 models by \citet{Castelli2003IAUS..210P.A20C}.  \label{fig:fig12_CCD_Ic}}
\end{figure*}

One of challenges in observing SN progenitors is the presence of possible companion or background/foreground stars. If a SN progenitor has a companion with a comparable luminosity or if massive background/foreground stars are present along the same line of sight, the detected object would contain information from both the progenitor and the contaminants. 

For comparison, we provide O/B-type ZAMS and evolved star models with $M_\mathrm{ZAMS}=9-120 M_\odot$, based on the models of \citet{Fierro2015PASP..127..428F}, as shown in Figures \ref{fig:fig4_HST_detect}-\ref{fig:fig6_NGRST_detect} and \ref{fig:fig8_Av_effect}-\ref{fig:fig10_CMD_Ic}. While such very massive companions would not be common for relatively low-mass stars like SN Ib/Ic progenitors, but \citet{Wellstein1999A&A...350..148W} show that SN Ib/Ic progenitors could have such a massive companion if the binary systems underwent conservative mass transfer. Additionally, for supernova progenitor detection in distant galaxies and for NGRST observations with a larger field of view than HST and JWST, considering physically unrelated foreground/background stars is important.

A $32-40\,M_{\odot}$ ZAMS star or an evolved star with an initial mass of $15-25\,M_{\odot}$ has a magnitude comparable to those of SN Ib/Ic progenitors with the fiducial mass-loss rates of in optical filters, as shown in Figure \ref{fig:fig4_HST_detect}. The optical colors of many SN Ib/Ic progenitors also overlap significantly with those of contaminants, as shown in the color-color diagrams (see left panels of Figures \ref{fig:fig11_CCD_Ib} and \ref{fig:fig12_CCD_Ic}). In such cases, it is necessary to wait until the SN becomes optically faint to confirm whether the detected candidate is indeed a progenitor or a contaminant.

Near-IR observations provide an additional advantage in distinguishing SN Ib/Ic progenitors from potential contaminants. Massive O/B-type stars have significantly lower near-IR brightness compared to their optical brightness, whereas SN Ib/Ic progenitors with optically thick winds are fairly bright in the near-IR. This characteristic enables near-IR observations to exclude the possibility of contaminants more effectively compared to optical observations. For example, with the F200W and F444W filters in JWST/NIRCam, only very massive stars, such as $85-120\,M_{\odot}$ ZAMS stars or evolved stars with an initial mass of $25-32\,M_{\odot}$, can have magnitudes comparable to those of SN Ib/Ic progenitors with the fiducial mass-loss rate, as shown in Figures \ref{fig:fig5_JWST_detect} and \ref{fig:fig6_NGRST_detect}. Such massive stars are relatively rare and are unlikely to be companions of SN Ib/Ic progenitors.

Moreover, SN Ib/Ic progenitors can be effectively distinguished from contaminants using near-IR colors. Figure \ref{fig:fig11_CCD_Ib} presents some color-color diagrams of SN Ib progenitor models. In the optical (left panels of the figure), SN Ib progenitors may have colors similar to those of contaminants. Even in near-IR (middle and right panels of the figure), some SN Ib progenitor models with low mass-loss rates exhibit weak free-free emission, leading to near-IR colors similar to those of O/B-type stars. However, SN Ib progenitors with mass-loss rates exceeding $\dot{M}_\mathrm{fid}$ have strong free-free emission, resulting in distinctive near-IR colors.

We present the corresponding color-color diagrams for SN Ic progenitors in Figure \ref{fig:fig12_CCD_Ic}. As previously discussed, the strong emission lines of SN Ic progenitors dramatically affect their colors, necessitating cautious use of broad-band colors in progenitor analysis. Consequently, most optical and near-IR color-color diagrams are challenging to use to distinguish SN Ic progenitors from contaminants. However, certain selected colors that include one or more near-IR filters, such as F070W-F150W and F150W-F200W in JWST/NIRCam or F062-F158 and F158-F184 in NGRST/WFI, can effectively distinguish SN Ic progenitors from contaminants (lower middle and lower right panels of Figure \ref{fig:fig12_CCD_Ic}). In general, colors that involve distant wavelength filters are more effective in distinguishing progenitors from contaminants, as the effect of free-free emission becomes more pronounced at longer wavelengths. Conversely, the colors between close wavelength filters are significantly influenced by emission lines, which distort the colors expected from blackbody radiation combined with free-free emission. See Appendix \ref{sec:appendixB} for more diverse color-color diagrams.

\section{Conclusions} \label{sec:conclusions}
In this study, we have investigated the detection limits and advantages of near-IR observations in the search of SN Ib/Ic progenitors using JWST and NGRST, compared to the capabilities of the HST. Our results demonstrate that near-IR observations provide significant advantages in identifying and analyzing SN Ib/Ic progenitors, primarily due to the reduced impact of extinction and the distinctiveness of near-IR colors.

First, near-IR instruments provide greater detection possibilities for SN Ib/Ic progenitors under significant dust extinction, where most SN Ib/Ic progenitors are typically found. JWST and NGRST would be more effective than HST in discovering SN Ib/Ic progenitors.

Second, we find that the mass-loss rates of SN Ib/Ic progenitors can be better constrained using near-IR magnitudes. Especially for giant SN Ib progenitors, optical magnitudes are insufficient to provide accurate constraints on their mass-loss rates. The near-IR filter magnitudes are highly influenced by free-free emission and this makes different mass-loss rate models more distinguishable.

Next, near-IR magnitudes and colors are helpful in distinguishing SN Ib/Ic progenitors from other stars. O/B-type stars, which are likely companions of SN Ib/Ic progenitors, have more rapidly decreasing flux at longer wavelengths compared to SN Ib/Ic progenitors. Only very massive stars ($>$85 and $>120$ $M_\odot$ ZAMS stars in $\sim2$ and $\sim4$ $\mu$m, respectively) could have magnitudes comparable to those of SN Ib/Ic progenitors with the fiducial mass-loss rate in near-IR, allowing us to exclude many possibilities of companions or background/foreground stars. While optical colors often overlap with those of main-sequence or evolved stars, some near-IR colors provide a more distinct separation, especially when considering filters with widely separated wavelengths.

Overall, our study suggests that JWST and NGRST can play a crucial role in advancing our understanding of SN Ib/Ic progenitors. Near-IR observations not only enhance detectability but also offer better constraints on progenitor properties and help in excluding contaminants. The detection of SN Ib/Ic progenitors will lead us to a better understanding of the late-time evolution of massive stars. Finally, we emphasize that exposure times of several to 20 minutes are not sufficient to detect the SN Ib/Ic progenitors at distances greater than 10 Mpc. Deeper observations using HST, JWST, and NGRST are needed to increase the number of directly detected SN Ib/Ic progenitors.

\section*{Acknowledgements}
This work has been supported by the National Research Foundation of Korea (NRF) grant (RS-2024-0035267 and 2019R1A2C201088514). The authors sincerely thank John Hillier for making the CMFGEN code publicly available.

\vspace{5mm}
\facilities{HST(WFC3), JWST(NIRCam), NGRST(WFI)}

\software{astropy \citep{astropy2013A&A...558A..33A, astropy2018AJ....156..123A, astropy2022ApJ...935..167A}, synphot \citep{synphot2018ascl.soft11001S}, stsynphot \citep{stsynphot2020ascl.soft10003S}, specutils \citep{specutils2019ascl.soft02012A}}

\bibliography{ref}{}
\bibliographystyle{aasjournal}

\appendix

\section{Optical and near-IR magnitudes of SN Ib/Ic progenitors with various mass-loss rates} \label{sec:appendixA}

\begin{deluxetable*}{c|cccc|cccccccc|ccccccc}
\rotate
\tablenum{4}
\tabletypesize{\scriptsize}
\tablecaption{Optical/Near-IR Magnitudes of SN Ib/Ic Progenitor Models\label{tab:AppendixA}}
\tablehead{
\nocolhead{}  & \multicolumn{4}{|c}{HST/WFC3}
& \multicolumn{8}{|c}{JWST/NIRCam} 
& \multicolumn{7}{|c}{NGRST/WFI}\\
 \colhead{Model} 
& \multicolumn{1}{|c}{F438W} & \colhead{F555W} & \colhead{F606W} & \colhead{F814W} 
& \multicolumn{1}{|c}{F070W} & \colhead{F090W} & \colhead{F115W} & \colhead{F150W} & \colhead{F200W} & \colhead{F277W} & \colhead{F356W} & \colhead{F444W} 
& \multicolumn{1}{|c}{F062} & \colhead{F087} & \colhead{F106} & \colhead{F129} & \colhead{F158} & \colhead{F184} & \colhead{F213}
}
\startdata
 \\
 & \multicolumn{19}{c}{$\dot{M}_\mathrm{fid}\times10$}\\
 \\
 \hline
HE2.91 & -5.29 & -5.17 & -5.10 & -4.79 & -4.93 & -4.66 & -4.46 & -4.07 & -3.80 & -3.31 & -2.97 & -2.93 & -5.05 & -4.69 & -4.55 & -4.28 & -4.02 & -3.88 & -3.70 \\ 
HE2.97 & -5.52 & -5.35 & -5.27 & -4.91 & -5.08 & -4.71 & -4.56 & -4.12 & -3.97 & -3.51 & -3.24 & -3.30 & -5.21 & -4.76 & -4.65 & -4.32 & -4.09 & -4.02 & -3.87 \\ 
HE4.09 & -5.80 & -5.76 & -5.73 & -5.41 & -5.59 & -5.19 & -5.39 & -4.82 & -4.96 & -4.47 & -4.26 & -4.50 & -5.70 & -5.21 & -5.45 & -4.99 & -4.84 & -4.94 & -4.89 \\ 
HE5.05 & -6.21 & -6.16 & -6.14 & -5.82 & -6.00 & -5.60 & -5.81 & -5.23 & -5.37 & -4.87 & -4.67 & -4.90 & -6.11 & -5.62 & -5.87 & -5.40 & -5.25 & -5.35 & -5.30 \\ 
CO5.18 & -6.69 & -6.47 & -6.38 & -6.11 & -6.29 & -5.89 & -5.63 & -5.22 & -5.12 & -4.56 & -4.20 & -4.07 & -6.37 & -5.90 & -5.77 & -5.44 & -5.20 & -5.23 & -4.91 \\ 
CO5.50 & -6.81 & -6.42 & -6.34 & -6.26 & -6.45 & -5.82 & -5.48 & -5.00 & -5.05 & -4.45 & -4.11 & -4.12 & -6.44 & -5.80 & -5.70 & -5.29 & -4.99 & -5.19 & -4.71 \\ 
CO6.17 & -6.81 & -6.40 & -6.31 & -6.23 & -6.40 & -5.80 & -5.44 & -4.94 & -4.97 & -4.38 & -4.04 & -4.03 & -6.40 & -5.78 & -5.66 & -5.24 & -4.93 & -5.11 & -4.63 \\ 
CO7.50 & -6.74 & -6.34 & -6.21 & -6.07 & -6.23 & -5.73 & -5.29 & -4.76 & -4.69 & -4.13 & -3.75 & -3.69 & -6.26 & -5.73 & -5.53 & -5.09 & -4.74 & -4.84 & -4.38 \\ 
CO9.09 & -7.17 & -6.96 & -6.75 & -6.35 & -6.45 & -6.38 & -5.71 & -5.20 & -4.95 & -4.46 & -4.05 & -3.90 & -6.68 & -6.37 & -6.09 & -5.55 & -5.16 & -5.07 & -4.76 \\ 
CO2.16 & -4.65 & -4.32 & -4.22 & -3.91 & -4.12 & -3.68 & -3.37 & -2.95 & -2.82 & -2.26 & -1.89 & -1.68 & -4.20 & -3.70 & -3.51 & -3.20 & -2.93 & -2.95 & -2.57 \\ 
CO3.93 & -4.35 & -4.22 & -4.15 & -3.91 & -4.07 & -3.71 & -3.48 & -3.12 & -2.97 & -2.47 & -2.14 & -2.02 & -4.13 & -3.73 & -3.59 & -3.30 & -3.10 & -3.06 & -2.81 \\ 
 \hline
 \\
 & \multicolumn{19}{c}{$\dot{M}_\mathrm{fid}\times5$}\\
 \\
 \hline
HE2.91 & -5.36 & -5.21 & -5.11 & -4.75 & -4.90 & -4.62 & -4.33 & -3.92 & -3.57 & -3.04 & -2.67 & -2.55 & -5.05 & -4.65 & -4.45 & -4.15 & -3.86 & -3.66 & -3.45 \\ 
HE2.97 & -5.51 & -5.30 & -5.19 & -4.76 & -4.95 & -4.58 & -4.32 & -3.84 & -3.59 & -3.08 & -2.76 & -2.80 & -5.12 & -4.62 & -4.44 & -4.08 & -3.79 & -3.67 & -3.48 \\ 
HE4.09 & -5.40 & -5.24 & -5.15 & -4.76 & -4.96 & -4.55 & -4.65 & -4.05 & -4.09 & -3.62 & -3.40 & -3.61 & -5.10 & -4.58 & -4.73 & -4.25 & -4.05 & -4.12 & -3.98 \\ 
HE5.05 & -5.76 & -5.61 & -5.52 & -5.15 & -5.35 & -4.95 & -5.07 & -4.45 & -4.52 & -4.03 & -3.81 & -4.02 & -5.47 & -4.98 & -5.15 & -4.66 & -4.46 & -4.55 & -4.40 \\ 
CO5.18 & -6.30 & -5.97 & -5.92 & -5.80 & -6.00 & -5.41 & -5.17 & -4.65 & -4.78 & -4.17 & -3.86 & -3.93 & -6.00 & -5.38 & -5.35 & -4.90 & -4.66 & -4.89 & -4.48 \\ 
CO5.50 & -6.06 & -5.64 & -5.55 & -5.46 & -5.63 & -5.05 & -4.69 & -4.18 & -4.19 & -3.62 & -3.28 & -3.31 & -5.63 & -5.04 & -4.91 & -4.48 & -4.16 & -4.32 & -3.87 \\ 
CO6.17 & -6.05 & -5.64 & -5.53 & -5.43 & -5.59 & -5.04 & -4.67 & -4.13 & -4.13 & -3.57 & -3.21 & -3.23 & -5.60 & -5.04 & -4.89 & -4.45 & -4.12 & -4.27 & -3.82 \\ 
CO7.50 & -6.03 & -5.68 & -5.53 & -5.32 & -5.46 & -5.07 & -4.60 & -4.03 & -3.93 & -3.41 & -3.01 & -2.97 & -5.55 & -5.08 & -4.85 & -4.38 & -4.00 & -4.06 & -3.66 \\ 
CO9.09 & -6.09 & -5.81 & -5.63 & -5.30 & -5.43 & -5.21 & -4.63 & -4.07 & -3.89 & -3.40 & -3.00 & -2.92 & -5.60 & -5.20 & -4.95 & -4.44 & -4.03 & -4.02 & -3.65 \\ 
CO2.16 & -3.95 & -3.60 & -3.51 & -3.33 & -3.52 & -3.01 & -2.68 & -2.24 & -2.19 & -1.62 & -1.26 & -1.13 & -3.54 & -3.03 & -2.85 & -2.50 & -2.22 & -2.32 & -1.91 \\ 
CO3.93 & -4.02 & -3.80 & -3.75 & -3.54 & -3.74 & -3.27 & -3.05 & -2.63 & -2.67 & -2.11 & -1.81 & -1.79 & -3.77 & -3.27 & -3.18 & -2.83 & -2.63 & -2.78 & -2.41 \\ 
 \hline
 \\
 & \multicolumn{19}{c}{$\dot{M}_\mathrm{fid}\times2$}\\
 \\
 \hline
HE2.91 & -5.40 & -5.23 & -5.13 & -4.75 & -4.91 & -4.61 & -4.28 & -3.85 & -3.41 & -2.84 & -2.40 & -2.18 & -5.06 & -4.66 & -4.41 & -4.09 & -3.77 & -3.53 & -3.29 \\ 
HE2.97 & -5.21 & -5.02 & -4.92 & -4.52 & -4.68 & -4.37 & -4.05 & -3.63 & -3.24 & -2.69 & -2.29 & -2.13 & -4.84 & -4.41 & -4.18 & -3.87 & -3.56 & -3.34 & -3.12 \\ 
HE4.09 & -5.02 & -4.73 & -4.57 & -4.11 & -4.32 & -3.92 & -3.80 & -3.25 & -3.07 & -2.61 & -2.33 & -2.43 & -4.50 & -3.97 & -3.91 & -3.49 & -3.22 & -3.20 & -2.89 \\ 
HE5.05 & -5.07 & -4.84 & -4.67 & -4.33 & -4.51 & -4.18 & -4.13 & -3.61 & -3.44 & -3.01 & -2.72 & -2.81 & -4.62 & -4.21 & -4.23 & -3.82 & -3.58 & -3.57 & -3.27 \\ 
CO5.18 & -5.56 & -5.22 & -5.12 & -4.98 & -5.14 & -4.65 & -4.35 & -3.77 & -3.78 & -3.22 & -2.88 & -2.94 & -5.17 & -4.67 & -4.54 & -4.07 & -3.76 & -3.88 & -3.53 \\ 
CO5.50 & -5.32 & -5.04 & -4.89 & -4.59 & -4.74 & -4.48 & -3.96 & -3.43 & -3.28 & -2.82 & -2.44 & -2.40 & -4.87 & -4.48 & -4.25 & -3.76 & -3.40 & -3.40 & -3.07 \\ 
CO6.17 & -5.32 & -5.08 & -4.91 & -4.55 & -4.69 & -4.53 & -3.95 & -3.43 & -3.25 & -2.79 & -2.42 & -2.37 & -4.87 & -4.52 & -4.29 & -3.76 & -3.40 & -3.38 & -3.05 \\ 
CO7.50 & -5.29 & -5.11 & -4.82 & -4.41 & -4.51 & -4.65 & -3.83 & -3.42 & -3.22 & -2.73 & -2.38 & -2.27 & -4.75 & -4.57 & -4.42 & -3.69 & -3.41 & -3.32 & -3.06 \\ 
CO9.09 & -5.28 & -5.11 & -4.70 & -4.34 & -4.47 & -4.57 & -3.80 & -3.44 & -3.29 & -2.76 & -2.44 & -2.28 & -4.63 & -4.51 & -4.41 & -3.68 & -3.45 & -3.37 & -3.16 \\ 
CO2.16 & -3.45 & -3.17 & -3.04 & -2.71 & -2.87 & -2.60 & -2.10 & -1.59 & -1.42 & -0.96 & -0.55 & -0.46 & -3.00 & -2.61 & -2.36 & -1.91 & -1.56 & -1.55 & -1.20 \\ 
CO3.93 & -3.54 & -3.21 & -3.12 & -2.98 & -3.15 & -2.64 & -2.39 & -1.83 & -1.86 & -1.31 & -0.99 & -1.08 & -3.17 & -2.66 & -2.56 & -2.11 & -1.82 & -1.95 & -1.63 \\ 
\enddata
\end{deluxetable*}

\begin{deluxetable*}{c|cccc|cccccccc|ccccccc}
\rotate
\tablenum{4}
\tabletypesize{\scriptsize}
\tablecaption{(Continued)}
\tablehead{ 
\nocolhead{}  & \multicolumn{4}{|c}{HST/WFC3}
& \multicolumn{8}{|c}{JWST/NIRCam} 
& \multicolumn{7}{|c}{NGRST/WFI}\\
 \colhead{Model} 
& \multicolumn{1}{|c}{F438W} & \colhead{F555W} & \colhead{F606W} & \colhead{F814W} 
& \multicolumn{1}{|c}{F070W} & \colhead{F090W} & \colhead{F115W} & \colhead{F150W} & \colhead{F200W} & \colhead{F277W} & \colhead{F356W} & \colhead{F444W} 
& \multicolumn{1}{|c}{F062} & \colhead{F087} & \colhead{F106} & \colhead{F129} & \colhead{F158} & \colhead{F184} & \colhead{F213}
}
\startdata
\\
 & \multicolumn{19}{c}{$\dot{M}_\mathrm{fid}\times1$}\\
 \\
 \hline
HE2.91 & -5.39 & -5.22 & -5.11 & -4.74 & -4.89 & -4.60 & -4.25 & -3.83 & -3.37 & -2.78 & -2.32 & -2.04 & -5.04 & -4.64 & -4.39 & -4.07 & -3.75 & -3.49 & -3.24 \\ 
HE2.97 & -5.25 & -5.04 & -4.93 & -4.50 & -4.67 & -4.35 & -3.99 & -3.54 & -3.09 & -2.52 & -2.08 & -1.85 & -4.85 & -4.39 & -4.13 & -3.80 & -3.47 & -3.22 & -2.97 \\ 
HE4.09 & -4.90 & -4.59 & -4.41 & -3.90 & -4.12 & -3.70 & -3.42 & -2.92 & -2.60 & -2.11 & -1.75 & -1.71 & -4.32 & -3.76 & -3.56 & -3.17 & -2.86 & -2.73 & -2.45 \\ 
HE5.05 & -4.74 & -4.45 & -4.26 & -3.88 & -4.05 & -3.71 & -3.58 & -3.14 & -2.92 & -2.49 & -2.17 & -2.13 & -4.19 & -3.75 & -3.68 & -3.33 & -3.10 & -3.03 & -2.77 \\ 
CO5.18 & -5.04 & -4.85 & -4.67 & -4.31 & -4.44 & -4.39 & -3.79 & -3.32 & -3.15 & -2.70 & -2.34 & -2.30 & -4.62 & -4.35 & -4.18 & -3.59 & -3.29 & -3.25 & -2.97 \\ 
CO5.50 & -4.72 & -4.51 & -4.15 & -3.83 & -3.97 & -4.07 & -3.34 & -3.05 & -2.88 & -2.38 & -2.09 & -1.95 & -4.09 & -3.99 & -3.90 & -3.24 & -3.05 & -2.93 & -2.77 \\ 
CO6.17 & -4.76 & -4.56 & -4.13 & -3.79 & -3.95 & -3.97 & -3.35 & -3.04 & -2.93 & -2.40 & -2.11 & -1.97 & -4.07 & -3.93 & -3.86 & -3.24 & -3.05 & -2.95 & -2.85 \\ 
CO7.50 & -4.66 & -4.61 & -4.14 & -3.69 & -3.86 & -3.60 & -3.25 & -2.90 & -2.92 & -2.30 & -1.99 & -1.88 & -4.07 & -3.68 & -3.55 & -3.13 & -2.92 & -2.87 & -2.88 \\ 
CO9.09 & -4.54 & -4.48 & -3.99 & -3.54 & -3.72 & -3.47 & -3.08 & -2.71 & -2.76 & -2.11 & -1.80 & -1.68 & -3.92 & -3.55 & -3.41 & -2.96 & -2.74 & -2.71 & -2.72 \\ 
CO2.16 & -2.84 & -2.63 & -2.31 & -2.02 & -2.13 & -2.26 & -1.52 & -1.20 & -0.99 & -0.54 & -0.23 & -0.08 & -2.25 & -2.19 & -2.07 & -1.40 & -1.19 & -1.06 & -0.89 \\ 
CO3.93 & -3.04 & -2.82 & -2.68 & -2.35 & -2.50 & -2.34 & -1.85 & -1.36 & -1.21 & -0.77 & -0.41 & -0.40 & -2.64 & -2.33 & -2.16 & -1.63 & -1.33 & -1.32 & -1.02 \\ 
 \hline
 \\
 & \multicolumn{19}{c}{$\dot{M}_\mathrm{fid}\times0.5$}\\
 \\
 \hline
HE2.91 & -5.43 & -5.25 & -5.14 & -4.75 & -4.91 & -4.62 & -4.26 & -3.82 & -3.35 & -2.75 & -2.27 & -1.95 & -5.07 & -4.66 & -4.40 & -4.08 & -3.75 & -3.47 & -3.22 \\ 
HE2.97 & -5.18 & -4.98 & -4.87 & -4.46 & -4.62 & -4.31 & -3.94 & -3.50 & -3.02 & -2.43 & -1.97 & -1.67 & -4.79 & -4.35 & -4.08 & -3.75 & -3.42 & -3.15 & -2.90 \\ 
HE4.09 & -5.05 & -4.73 & -4.54 & -3.99 & -4.22 & -3.79 & -3.41 & -2.92 & -2.52 & -2.00 & -1.61 & -1.45 & -4.44 & -3.85 & -3.56 & -3.19 & -2.85 & -2.65 & -2.37 \\ 
HE5.05 & -4.54 & -4.24 & -4.04 & -3.59 & -3.78 & -3.40 & -3.17 & -2.75 & -2.52 & -2.14 & -1.83 & -1.76 & -3.96 & -3.44 & -3.30 & -2.96 & -2.70 & -2.62 & -2.37 \\ 
CO5.18 & -4.50 & -4.31 & -3.85 & -3.52 & -3.67 & -3.65 & -3.16 & -2.87 & -2.78 & -2.25 & -1.98 & -1.87 & -3.79 & -3.62 & -3.57 & -3.05 & -2.87 & -2.75 & -2.73 \\ 
CO5.50 & -3.96 & -3.94 & -3.58 & -3.11 & -3.28 & -2.89 & -2.71 & -2.39 & -2.35 & -1.82 & -1.54 & -1.45 & -3.50 & -3.01 & -2.86 & -2.60 & -2.41 & -2.34 & -2.31 \\ 
CO6.17 & -3.82 & -4.03 & -3.75 & -3.13 & -3.29 & -2.83 & -2.78 & -2.40 & -2.37 & -1.87 & -1.54 & -1.45 & -3.66 & -3.00 & -2.87 & -2.67 & -2.45 & -2.43 & -2.28 \\ 
CO7.50 & -3.28 & -3.42 & -3.19 & -2.88 & -3.03 & -2.60 & -2.67 & -2.28 & -2.11 & -1.87 & -1.60 & -1.53 & -3.12 & -2.77 & -2.70 & -2.56 & -2.33 & -2.35 & -1.93 \\ 
CO9.09 & -3.16 & -3.22 & -2.99 & -2.77 & -2.92 & -2.51 & -2.57 & -2.18 & -2.00 & -1.75 & -1.49 & -1.42 & -2.93 & -2.67 & -2.59 & -2.45 & -2.22 & -2.22 & -1.82 \\ 
CO2.16 & -2.00 & -2.07 & -1.79 & -1.31 & -1.47 & -1.09 & -0.96 & -0.65 & -0.54 & -0.08 & 0.20 & 0.31 & -1.72 & -1.21 & -1.08 & -0.87 & -0.66 & -0.58 & -0.46 \\ 
CO3.93 & -2.50 & -2.29 & -1.91 & -1.61 & -1.75 & -1.81 & -1.24 & -0.97 & -0.79 & -0.32 & -0.05 & 0.06 & -1.86 & -1.74 & -1.69 & -1.14 & -0.95 & -0.80 & -0.72 \\ 
 \hline
 \\
 & \multicolumn{19}{c}{$\dot{M}_\mathrm{fid}\times0.1$}\\
 \\
 \hline
HE2.91 & -5.41 & -5.23 & -5.13 & -4.74 & -4.89 & -4.61 & -4.25 & -3.82 & -3.35 & -2.76 & -2.28 & -1.92 & -5.05 & -4.65 & -4.39 & -4.07 & -3.75 & -3.47 & -3.23 \\ 
HE2.97 & -5.19 & -4.99 & -4.87 & -4.45 & -4.62 & -4.30 & -3.92 & -3.47 & -2.96 & -2.36 & -1.86 & -1.51 & -4.79 & -4.35 & -4.07 & -3.73 & -3.39 & -3.10 & -2.84 \\ 
HE4.09 & -5.02 & -4.69 & -4.51 & -3.92 & -4.16 & -3.69 & -3.22 & -2.69 & -2.14 & -1.49 & -0.98 & -0.67 & -4.40 & -3.77 & -3.40 & -3.00 & -2.60 & -2.30 & -1.99 \\ 
HE5.05 & -4.49 & -4.16 & -3.96 & -3.37 & -3.62 & -3.13 & -2.68 & -2.14 & -1.68 & -1.09 & -0.63 & -0.45 & -3.86 & -3.20 & -2.86 & -2.44 & -2.06 & -1.82 & -1.49 \\ 
CO5.18 & -2.73 & -2.90 & -2.70 & -1.93 & -2.14 & -1.65 & -1.62 & -1.17 & -1.00 & -0.71 & -0.41 & -0.38 & -2.59 & -1.79 & -1.69 & -1.46 & -1.21 & -1.23 & -0.76 \\ 
CO5.50 & -2.20 & -2.07 & -1.84 & -1.43 & -1.62 & -1.17 & -1.12 & -0.69 & -0.47 & -0.20 & 0.07 & 0.14 & -1.75 & -1.31 & -1.17 & -0.99 & -0.72 & -0.69 & -0.30 \\ 
CO6.17 & -2.02 & -1.91 & -1.67 & -1.33 & -1.50 & -1.07 & -1.05 & -0.63 & -0.41 & -0.14 & 0.14 & 0.22 & -1.59 & -1.21 & -1.09 & -0.92 & -0.66 & -0.63 & -0.24 \\ 
CO7.50 & -1.61 & -1.47 & -1.28 & -0.97 & -1.11 & -0.78 & -0.62 & -0.33 & -0.10 & 0.16 & 0.39 & 0.43 & -1.21 & -0.87 & -0.69 & -0.51 & -0.33 & -0.24 & 0.01 \\ 
CO9.09 & -1.52 & -1.34 & -1.18 & -0.84 & -0.98 & -0.67 & -0.48 & -0.21 & 0.04 & 0.30 & 0.51 & 0.54 & -1.10 & -0.75 & -0.56 & -0.37 & -0.19 & -0.07 & 0.12 \\ 
CO2.16 & 0.47 & 0.53 & 0.69 & 0.93 & 0.81 & 1.08 & 1.17 & 1.41 & 1.58 & 1.79 & 1.98 & 1.98 & 0.74 & 1.00 & 1.13 & 1.26 & 1.41 & 1.47 & 1.67 \\ 
CO3.93 & -1.21 & -1.29 & -1.10 & -0.23 & -0.45 & 0.04 & 0.22 & 0.65 & 0.78 & 1.21 & 1.51 & 1.55 & -0.98 & -0.07 & 0.11 & 0.38 & 0.64 & 0.66 & 0.94 \\ 
\enddata
\end{deluxetable*}

In Table \ref{tab:AppendixA}, we present the absolute AB magnitudes of SN Ib/Ic progenitor models with various mass-loss rates in the HST/WFC3, JWST/NIRCam, and NGRST/WFI broad-band filters.

\clearpage

\section{Color-color diagrams of SN Ib/Ic progenitors in HST, JWST, and NGRST filters}\label{sec:appendixB}

\begin{figure*}[!htbp]
\includegraphics[width=\textwidth]{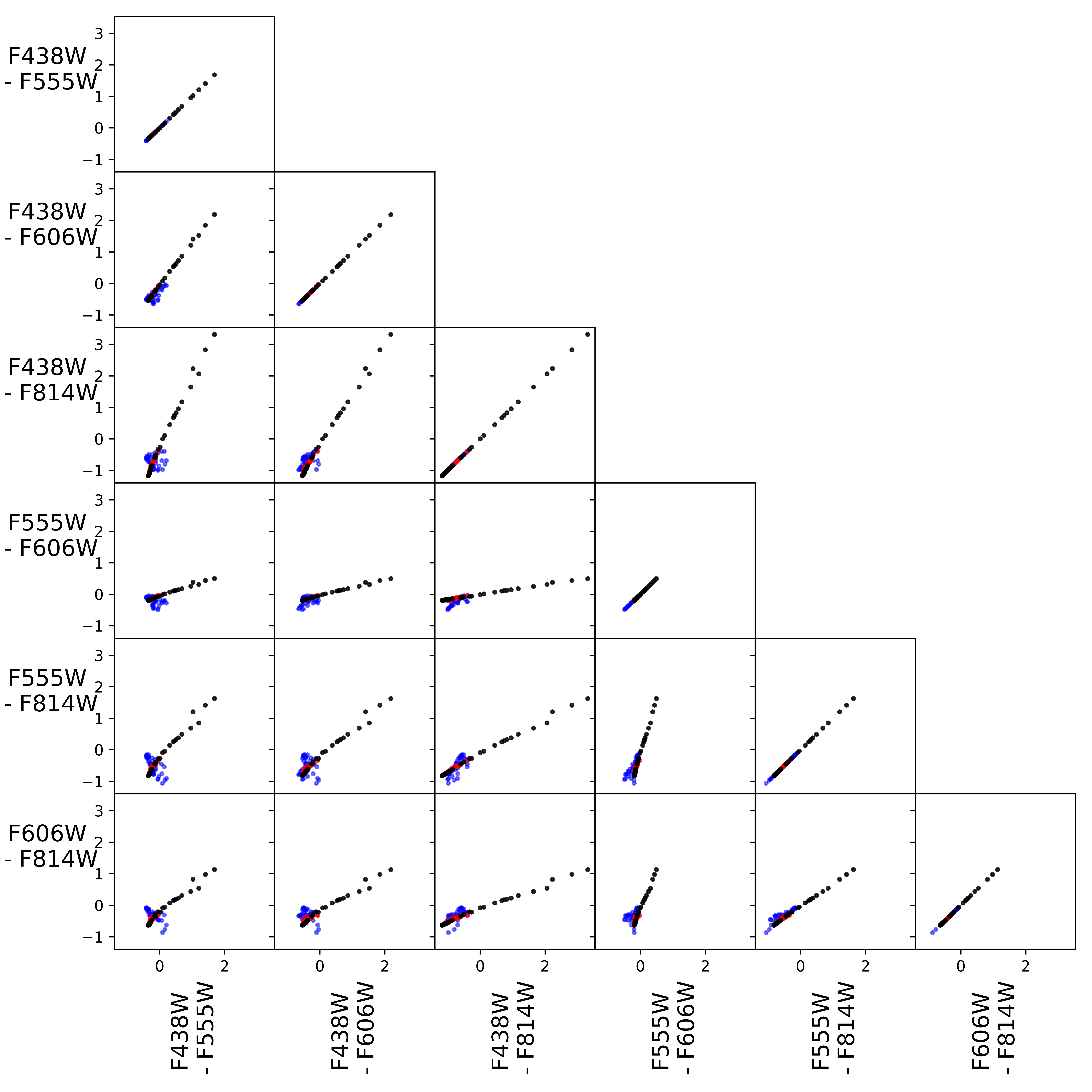}
\centering
\caption{Color-Color diagrams using various broad-band filters of HST/WFI. Blue and red markers denote SN Ib and Ic progenitor models, respectively. Possible companion or background/foreground stars (O/B-type stars, A to M type main-sequence stars, giant and supergiant stars) are presented with black markers. \label{fig:fig13_HST_ccd}}
\end{figure*}

In Figure \ref{fig:fig13_HST_ccd}, \ref{fig:fig14_1_JWST_ccd}, and \ref{fig:fig15_1_NGRST_ccd}, we present color-color diagrams of all combination of broad-band filters in HST/WFC3, JWST/NIRCam, and NGRST/WFI, respectively.

\begin{figure*}[!htbp]
\includegraphics[width=\textwidth]{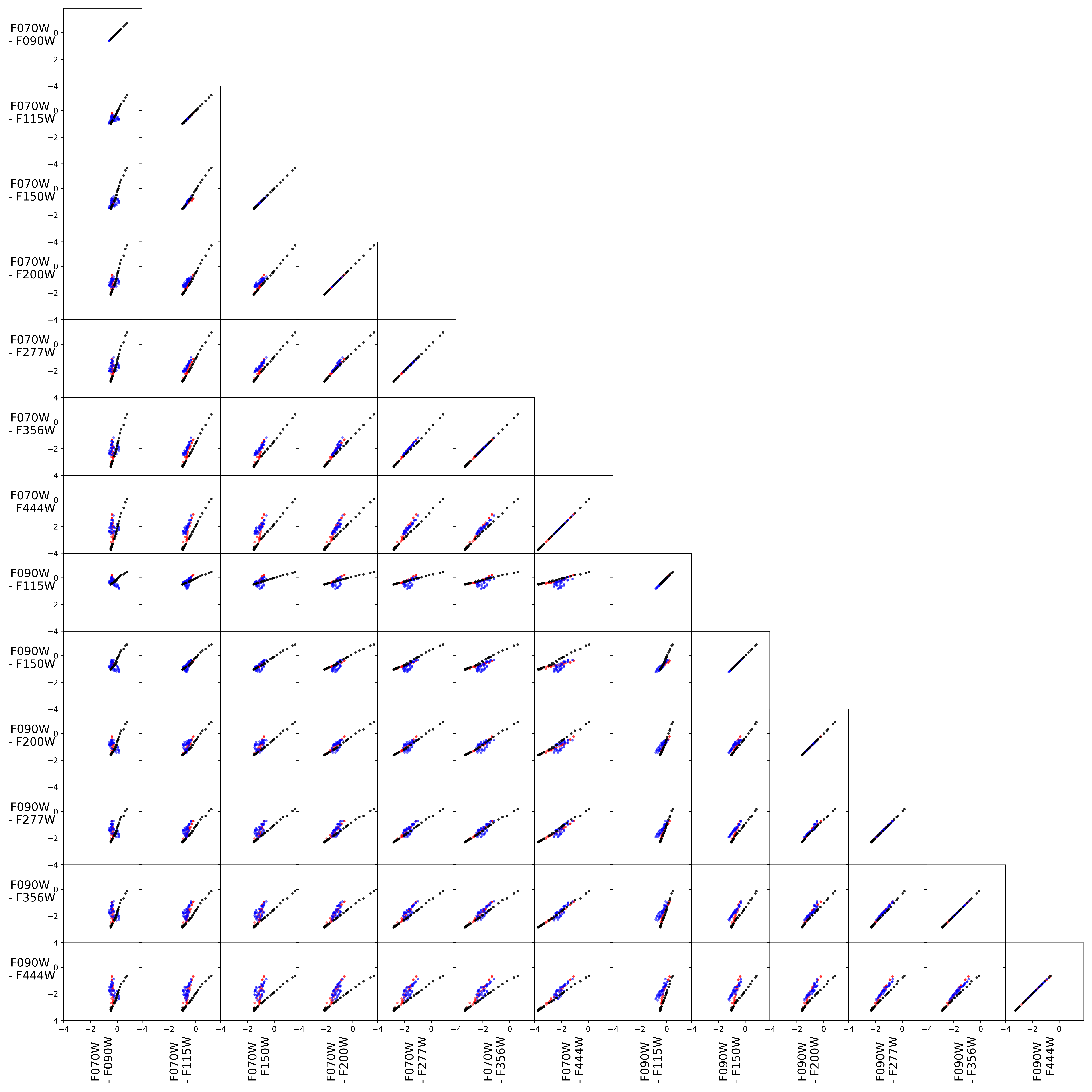}
\centering
\caption{Color-Color diagrams using various broad-band filters of JWST/NIRCam. Blue and red markers denote SN Ib and Ic progenitor models, respectively. The meaning of markers are the same with Figure \ref{fig:fig13_HST_ccd}. \label{fig:fig14_1_JWST_ccd}}
\end{figure*}

\begin{figure*}[!htbp]
\addtocounter{figure}{-1}
\renewcommand{\thefigure}{\arabic{figure} (continued)}
\centering
\includegraphics[width=\textwidth]{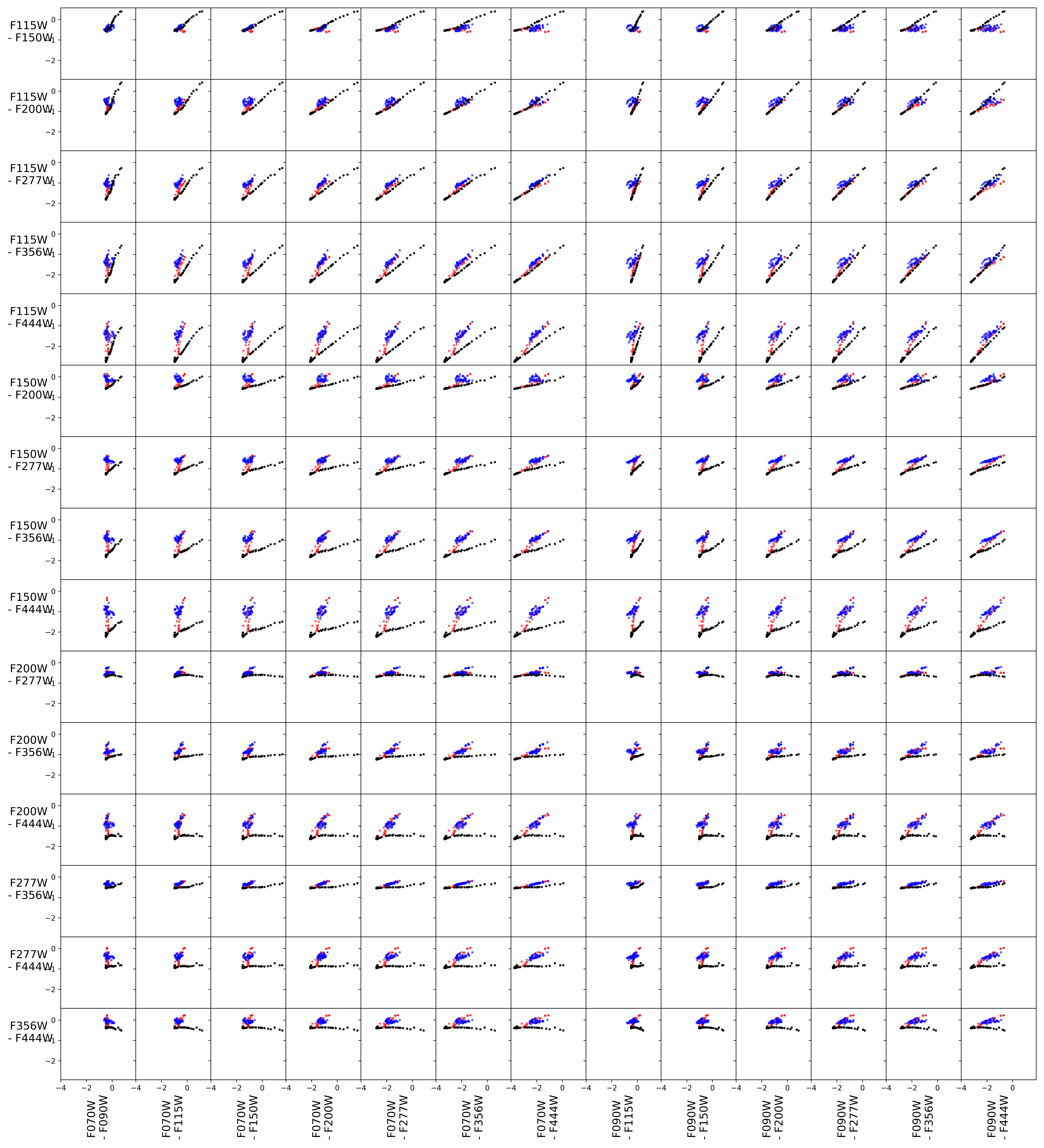}
\caption{Color-Color diagrams with various JWST/NIRCam broad-band filters. \label{fig:fig14_2_JWST_ccd}}
\end{figure*}

\begin{figure*}[!htbp]
\addtocounter{figure}{-1}
\renewcommand{\thefigure}{\arabic{figure} (continued)}
\centering
\includegraphics[width=\textwidth]{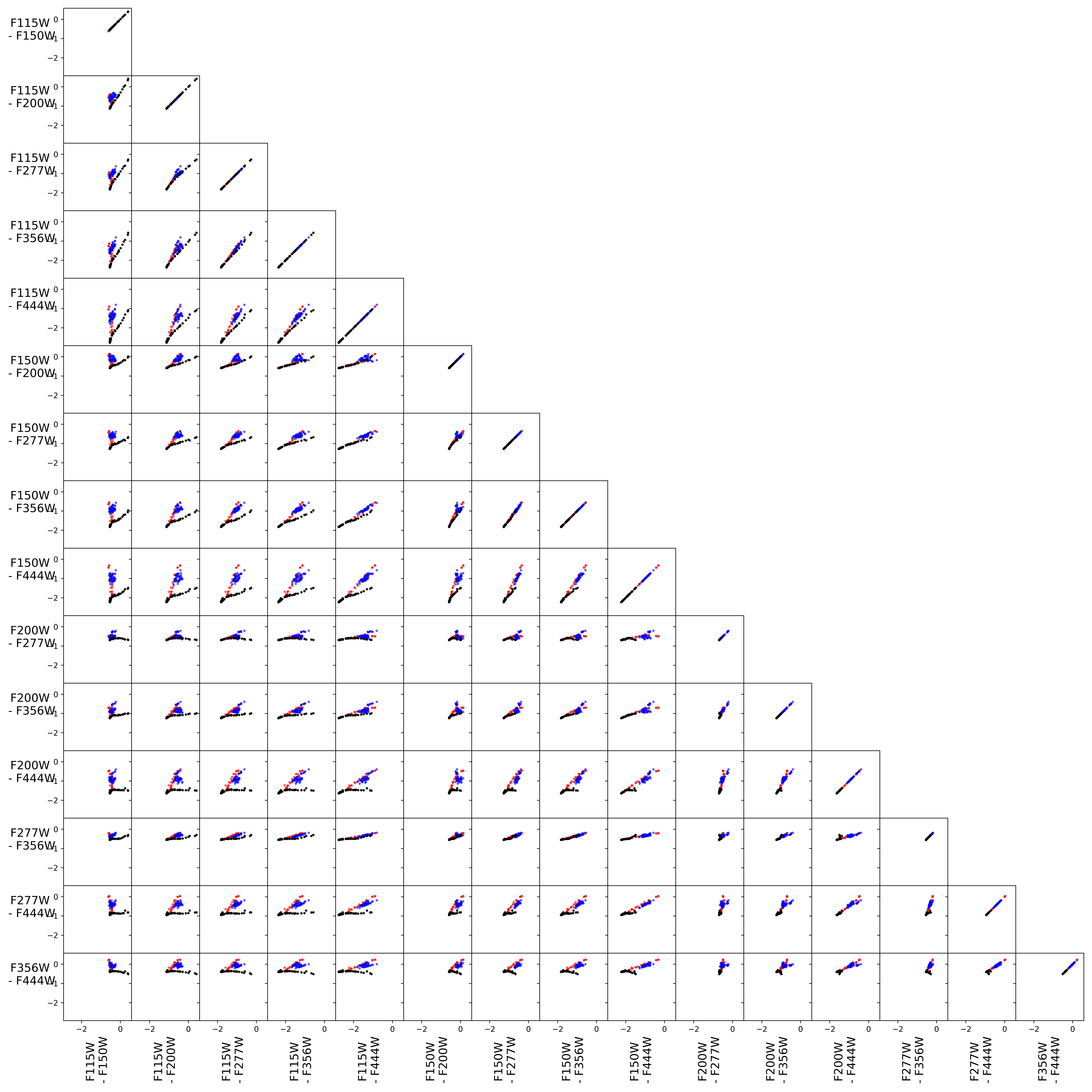}
\caption{Color-Color diagrams with various JWST/NIRCam broad-band filters. \label{fig:fig14_3_JWST_ccd}}
\end{figure*}

\begin{figure*}[!htbp]
\includegraphics[width=\textwidth]{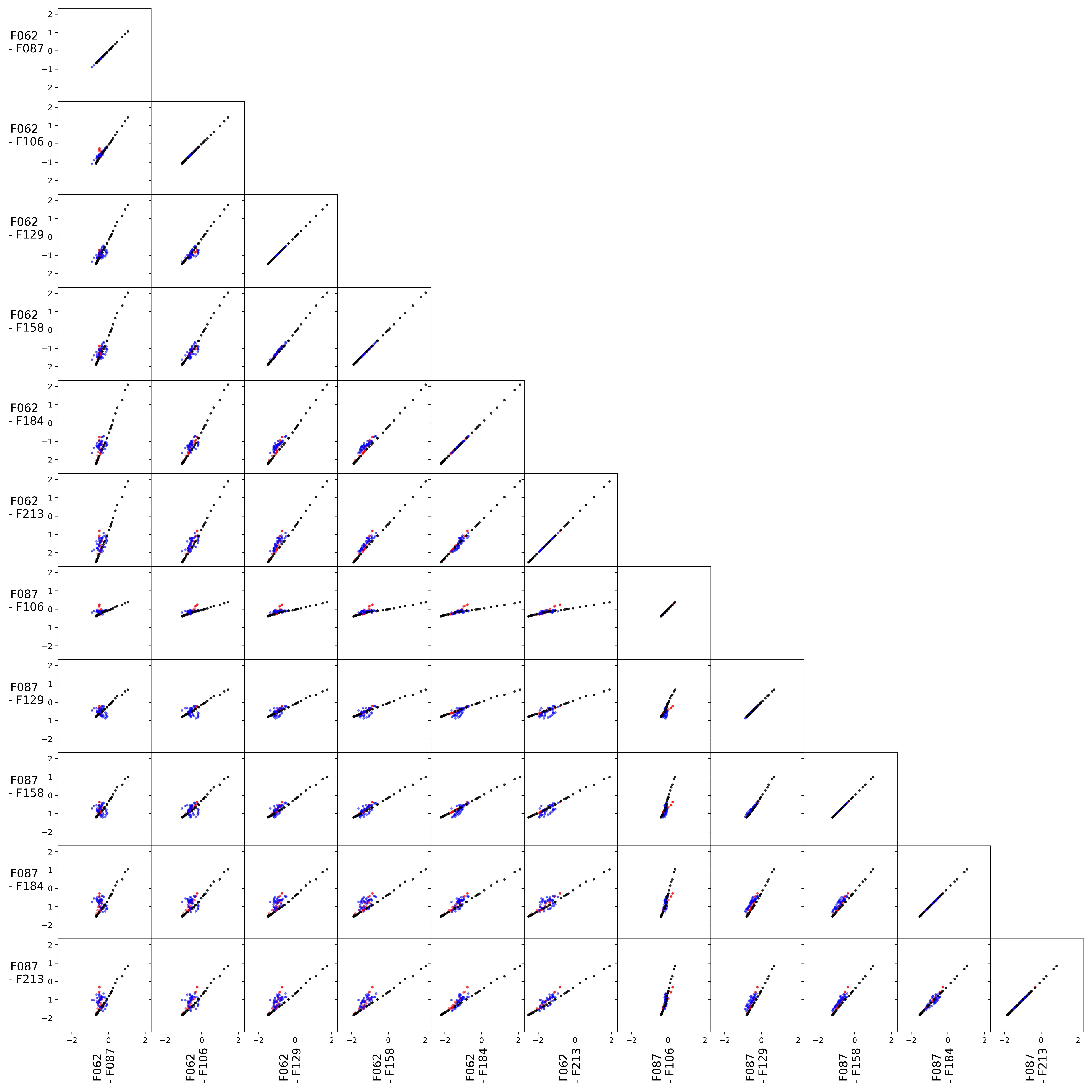}
\centering
\caption{Color-Color diagrams with various NGRST/WFI broad-band filters. Blue and red markers denote SN Ib and Ic progenitor models, respectively.  The meaning of markers are the same with Figure \ref{fig:fig13_HST_ccd}. \label{fig:fig15_1_NGRST_ccd}}
\end{figure*}

\begin{figure*}[!htbp]
\addtocounter{figure}{-1}
\renewcommand{\thefigure}{\arabic{figure} (continued)}
\centering
\includegraphics[width=\textwidth]{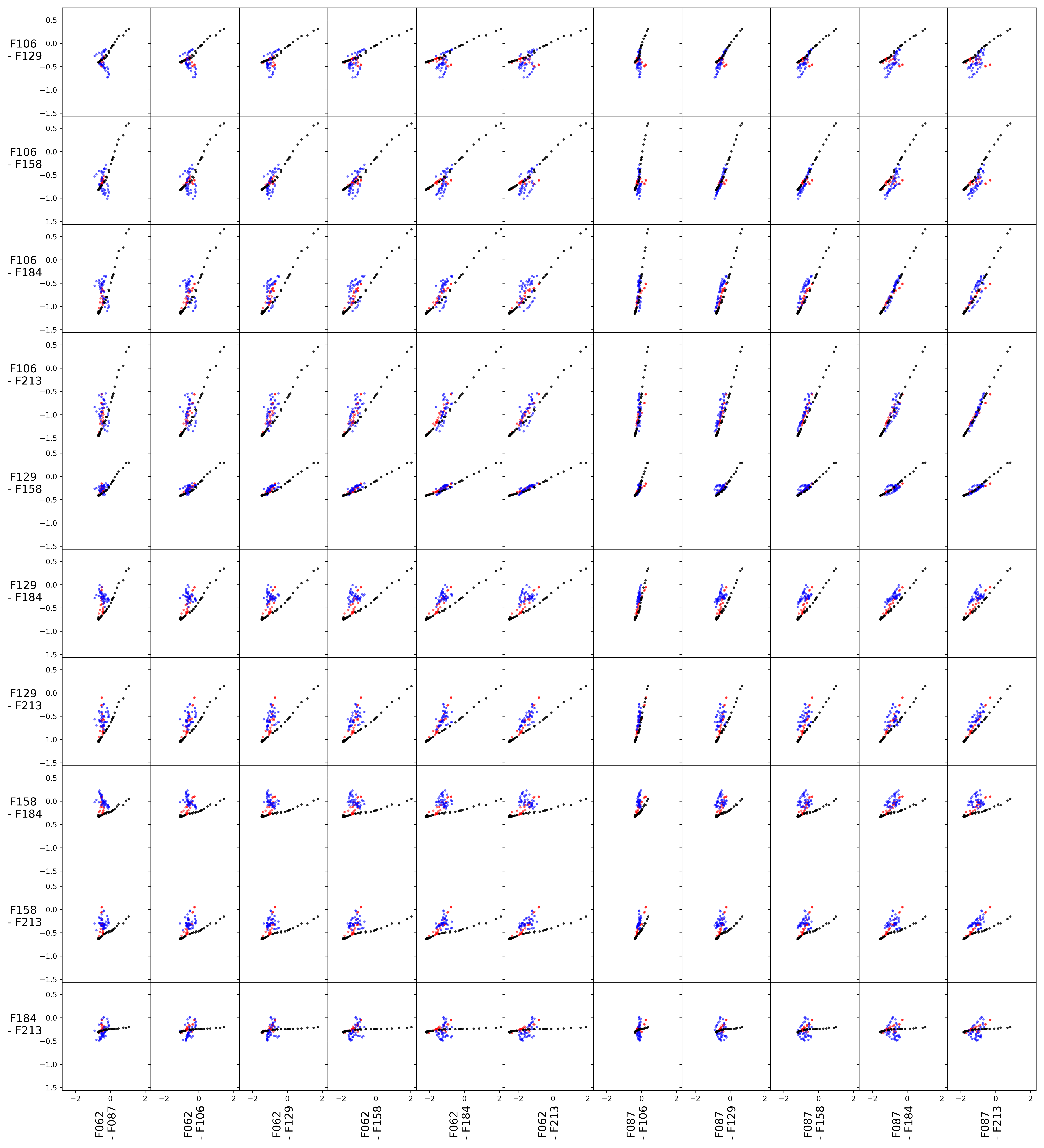}
\caption{Color-Color diagrams with various NGRST/WFI broad-band filters. \label{fig:fig15_2_NGRST_ccd}}
\end{figure*}

\begin{figure*}[!htbp]
\addtocounter{figure}{-1}
\renewcommand{\thefigure}{\arabic{figure} (continued)}
\centering
\includegraphics[width=\textwidth]{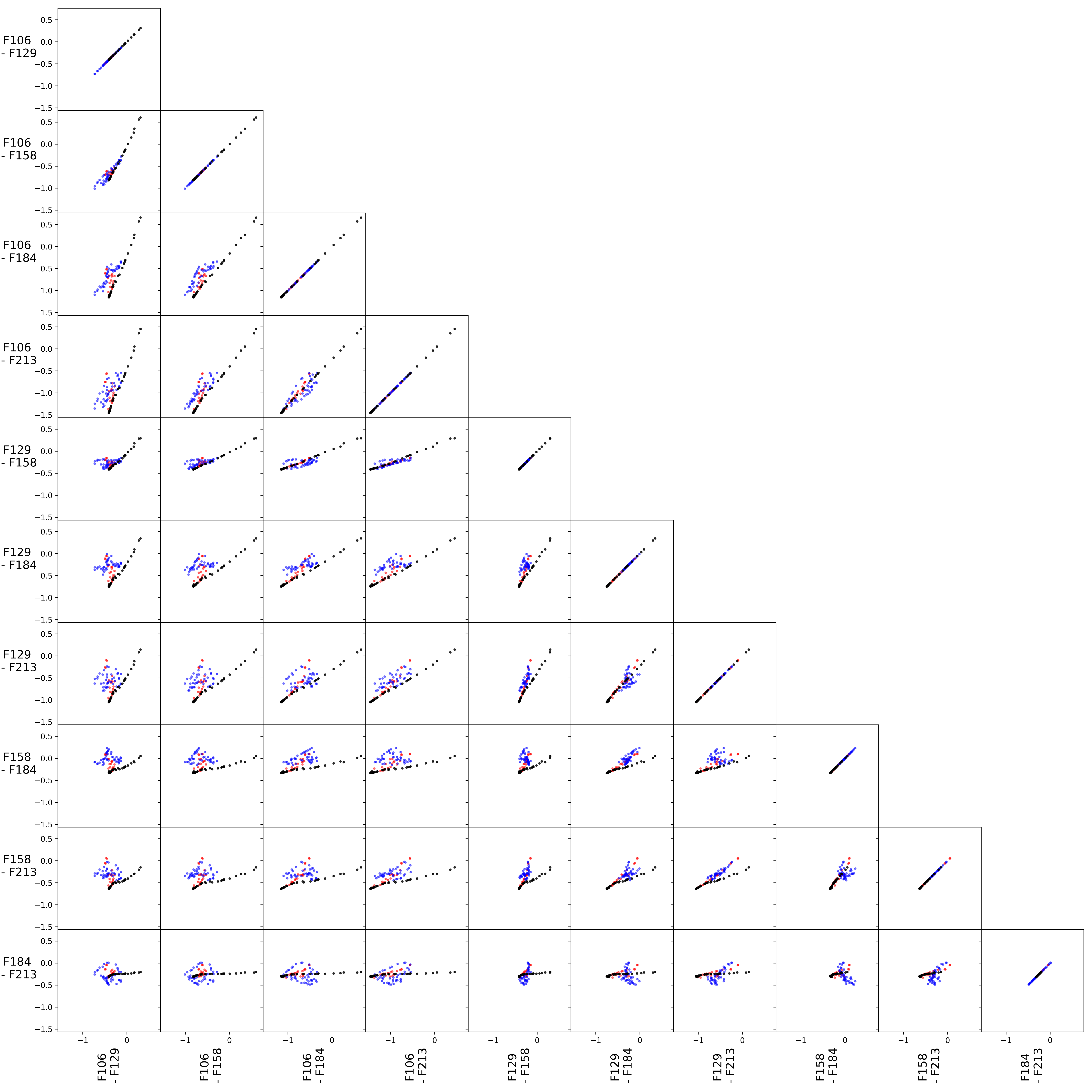}
\caption{Color-Color diagrams with various NGRST/WFI broad-band filters. \label{fig:fig15_3_NGRST_ccd}}
\end{figure*}

\end{document}